\documentclass[12pt,a4paper,final]{iopart}
\usepackage{iopams}
\expandafter\let\csname equation*\endcsname\relax
\expandafter\let\csname endequation*\endcsname\relax
\usepackage{amsmath,amssymb}

\usepackage{graphicx}
\usepackage{mathrsfs}
\usepackage{url}
\usepackage[usenames]{color}
\usepackage{ulem}
\usepackage{cases}
\newcommand{\domega}{\delta\!\!\;\omega}
\newcommand{\dk}{\delta\!\!\;k}
\newcommand{\hdomega}{\delta\!\!\;\hat{\omega}}
\newcommand{\pont}{{\,^\ast\!}RR}

\def\dd{\mathrm{d}}

\def\Mpl{M_{\rm Pl}}
\def\GeV{{\rm GeV}}
\def\eV{{\rm eV}}
\def\km{{\rm km}}
\def\kpc{{\rm kpc}}
\def\erf{{\rm erf}}

\begin{document}

\title{Resonant gravitational waves in dynamical Chern--Simons--axion gravity}

\author{%
Tomohiro~Fujita$^{1}$,
Ippei~Obata$^{2,3}$,
Takahiro~Tanaka$^{2,4}$,
Kei~Yamada$^{2}$
}
\address{$^{1}$~Institute for Cosmic Ray Research, University of Tokyo, Kashiwa 277-8582, Japan.}
\address{$^{2}$~Department of Physics, Kyoto University, Kyoto 606-8502, Japan.}
\address{$^{3}$~Max-Planck-Institut f{\"u}r Astrophysik, Karl-Schwarzschild-Str. 1, 85741 Garching, Germany.}
\address{$^{4}$~Center for Gravitational Physics, Yukawa Institute for Theoretical Physics, Kyoto University, Kyoto 606-8502, Japan.}

\ead{k.yamada@tap.scphys.kyoto-u.ac.jp}

\date{\today}

\begin{abstract}
In this paper, we consider dynamical Chern--Simons gravity with the identification of the scalar field coupled though the Pontryagin density with the axion dark matter, and we discuss the effects of the parametric resonance on gravitational waves (GWs). When we consider GWs in a coherently oscillating axion cloud, we confirm that significant resonant amplification of GWs occurs in a narrow frequency band, and the amplification is restricted to the late epoch after the passage of the incident waves. We also identify the condition that an axion cloud spontaneously emits GWs. Once we take into account the randomness of the spatial phase distribution of the axion oscillations, we find that the amplification is suppressed compared with the coherent case, but significant amplification of GWs can still occur. We also examine whether or not the amplification of GWs is possible in the present universe, taking into account the history of the universe. We find that resonant amplification is difficult to be tested from GW observations in the standard scenario of the axion DM model, in which the axion is the dominant component of DM. However, there is some parameter window in which the resonant amplification of GWs might be observed, if the axion is subdominant component of DM, and the axion cloud formation is delayed until the Hubble rate becomes much smaller than the axion mass.
\end{abstract}

\maketitle

\section{Introduction}

The gravitational waves (GWs) provide us with opportunities to perform new tests of general relativity (GR) in the strong gravity regime. Regarding observing runs O1 and O2 by LIGO and Virgo collaborations, ten binary black hole (BH) mergers and one neutron star (NS) binary merger are summarized in the catalog GWTC--1~\cite{LIGOScientific:2018mvr}. A few interesting events in O3 have already been reported~\cite{Abbott:2020uma,LIGOScientific:2020stg,Abbott:2020khf} and the summary paper covering the whole O3 will come soon. Also, KAGRA has started observing run from the end of the last February. Testing GR using the O1/O2 GW data has been done by several authors, and they reported no significant deviation from GR (for instance, see Refs.~\cite{TheLIGOScientific:2016src, Yunes:2016jcc, Abbott:2018lct,LIGOScientific:2019fpa,Nair:2019iur,Uchikata:2019frs,Yamada:2019zrb,Saffer:2020xsw,Carson:2020ter,Yamada:2020zvt}). However, there remains the possibility that some deviation might be found by performing an analysis tuned for some particular class of modifications to GR. 

One possibility that has attracted some attention in high energy physics is gravitational parity--violation such as represented by dynamical Chern--Simons (dCS) gravity~\cite{Jackiw:2003pm,Alexander:2009tp}. This theory modifies GR by adding a dynamical pseudoscalar field coupled non--minimally to curvature via the Pontryagin density. The theory is also motivated from the anomaly cancellation in heterotic string theory upon 4--dimensional compactification and a low--energy expansion~\cite{Green:1984sg} and from loop quantum gravity upon the promotion of the Barbero--Immirzi parameter to a field in the presence of matter~\cite{Taveras:2008yf,Calcagni:2009xz}. We treat this theory as an effective model, valid only at sufficiently low--energies relative to some cut--off scale. The magnitude of dCS deformation from GR is controlled by the dimensionful coupling parameter $\ell_{\rm dCS}$. Observations of Lense--Thirring precession obtained by the Gravity Probe B experiment~\cite{Everitt:2011hp} and the LAGEOS/LARES satellites~\cite{Ciufolini:2016ntr} place an approximate constraint as $\ell_{\rm dCS} \lesssim {\cal{O}}(10^{8} \, \km)$~\cite{AliHaimoud:2011fw,Nakamura:2018yaw}. In the future, the observation of GWs emitted by spinning black holes will lead to an eight--order of magnitude improvement on these constraints~\cite{Loutrel:2018rxs}, once we detect GWs that are sufficiently strong to break degeneracies between the spins of the objects and the dCS deformation. Very recently, a more stringent constraint, $\ell_{\rm dCS} \lesssim 10 \, \rm{km}$, has been obtained by Silva {\it et al.}~\cite{Silva:2020acr} by using the results of {\it Neutron Star Interior Composition Explorer} ({\it NICER}) measuring the mass and equatorial radius of the isolated neutron star PSR~J0030+0451.%
\footnote{To evaluate the momentum of inertia of an isolated neutron star they have assumed GR, which may affect their constraint.}
Although this constraint is derived under the assumption that the scalar field is massless, it can apply to a massive scalar field as long as its Compton wavelength is much longer than $10 \, \rm{km}$. Therefore, the constraint also applies to dCS--axion gravity, in which the scalar field is identified with an axions/axion--like particle (hereafter we simply refer to them as the axions), which we discuss in this paper.

The axions predicted by string theory are known to acquire the mass by the quantum non--perturbative effect such as the instanton effect, and the spectrum of possible mass is logarithmically broad ~\cite{Svrcek:2006yi,Arvanitaki:2009fg}. Especially, the axion may be a component of dark matter (DM), behaving as a non--relativistic fluid after it starts coherent oscillations in the history of the universe~\cite{Marsh:2015xka}. Therefore, there are many studies to probe the axion DM (for example, Refs.~\cite{Hlozek:2017zzf, Marsh:2010wq, Khmelnitsky:2013lxt, Porayko:2014rfa, Aoki:2017ehb, Blas:2016ddr, Arvanitaki:2010sy, Yoshino:2015nsa, Abel:2017rtm, Fujita:2018zaj,Nagano:2019rbw}). Furthermore,  owing to the gravitational Chern--Simons coupling and the axion's oscillatory feature, the possibility of significant amplification of propagating GWs via the parametric resonance mechanism has recently been pointed out~\cite{Yoshida:2017cjl,Chu:2020iil,Jung:2020aem}. This phenomenon might be important to give a constraint on dCS--axion gravity as well as to search for the axion DM.

In this paper, therefore, we investigate the effects of the parametric resonance on GWs toward seeking the signature of modification in GW signals. The rest of this paper is organized as follows. Section~\ref{Sec:Basics_dCS} briefly reviews dCS gravity and a simple description of the parametric resonance in the coherent axion case. In Sec.~\ref{Sec:coherentcase}, focusing on the case of the coherent axion, we scrutinize how the resonant amplification of GWs occurs and when it becomes significant. We also discuss the possibility that the axion cloud spontaneously emits GWs. Section~\ref{Sec:the effect of spatial distribution} investigates the effect of the loss of coherence on the resonant amplification by taking into account the spatial variation of the axion amplitude and phase corresponding to the velocity dispersion. In Sec.~\ref{Sec:Condition for the GW resonance}, we discuss how much amplification of GWs is allowed in the present universe by taking into account the cosmic expansion and the backreaction of GW emission upon the axion cloud and explore a scenario in which significant amplification of GWs might be observed. Finally, Sec.~\ref{Sec:Summary} is devoted to summary and discussion. We adopt the conventions of Ref.~\cite{misner1973gravitation}, in particular for the signature of the metric, Riemann, and Einstein tensors. Throughout this paper we use natural units in which $\hbar = 1 = c$.

\section{Equation of motion of GWs in dCS--axion gravity}
\label{Sec:Basics_dCS}

In this section, we describe the basics of dCS gravity. The action is given by~\cite{Alexander:2009tp}
\begin{align}
  \label{Action}
  S \equiv \int d^4 x \sqrt{-g} \left[ \frac{\Mpl^2}{2} R + \frac{\Mpl \ell_{\rm dCS}^2}{4 \sqrt{2}} \phi \, {^\ast\!} R R - \frac12 \left( \nabla_{\mu} \phi \nabla^{\mu} \phi + 2 V(\phi) \right)+ \mathcal{L}_{\rm mat} \right] ,
  \end{align}
where $\Mpl \equiv (8 \pi G)^{- 1/2} \approx 2 \times 10^{18} \, \GeV$ is the reduced Planck mass, $G$ is the Newtonian constant, $g$ denotes the determinant of the metric $g_{\mu \nu}$, $R$ is the Ricci scalar, $\phi$ is the (pseudo) scalar field, $V(\phi)$ is its potential, $\nabla_{\alpha}$ denotes the covariant differentiation, $\mathcal{L}_{\rm mat}$ is the matter Lagrangian density, and $\ell_{\rm dCS}$ is a coupling parameter. In the following, we take the convention that $\phi$ and $\ell_{\rm dCS}$ have mass dimension 1 and -1, respectively.

The Pontryagin density $\pont$ is defined by~\cite{Alexander:2009tp}
\begin{align}
  \label{Pont}
  {^\ast\!} R R \equiv {^\ast\!} R^{\mu \nu \rho \sigma} R_{\nu \mu \rho \sigma} , \qquad
  {^\ast\!} R^{\mu \nu \rho \sigma} \equiv \frac{1}{2} \epsilon^{\rho \sigma \alpha \beta} R^{\mu \nu}_{~~ \alpha \beta} ,
\end{align}
where $\epsilon^{\mu \nu \rho \sigma}$ is the Levi--Civita tensor with $\epsilon^{0123}=-1/\sqrt{-g}$. If the (pseudo) scalar field $\phi$ is constant, dCS gravity identically reduces  to GR, because the Pontryagin density term in the action is the total divergence of the topological current~\cite{Jackiw:2003pm}, and therefore it does not contribute to the field equations. 

The field equations of dCS gravity are obtained by varying the action~\eqref{Action} with respect to the metric $g_{\mu \nu}$ and the scalar field $\phi$~\cite{Alexander:2009tp}:
\begin{align}
  \label{eq;metric}
G_{\mu \nu} + \frac{\sqrt{2} \ell_{\rm dCS}^2}{\Mpl} C_{\mu \nu}
  &= \frac{1}{\Mpl^2} \left( T^{\rm mat}_{\mu \nu} + T^{\phi}_{\mu \nu} \right) , \\
  \label{eq;scalar}
  \Box \phi
  &= \frac{d V}{d \phi} - \frac{\ell_{\rm dCS}^2 \Mpl}{4 \sqrt{2}} {^\ast\!} R R ,
\end{align}
where $G_{\mu \nu}$ is the Einstein tensor and $T^{\rm mat}_{\mu \nu}$ is the stress--energy tensor of the matter field. The d'Alembertian operator is here denoted by $\Box \equiv \nabla_{\alpha} \nabla^{\alpha}$. The $C$--tensor and the stress--energy tensor for the scalar field are defined by 
\begin{align}
  C^{\mu \nu} &\equiv \left( \nabla_{\sigma} \phi \right) \epsilon^{\sigma \delta \alpha ( \mu} \nabla_{\alpha} R^{\nu )}{}_{\delta} + \left( \nabla_{\sigma} \nabla_{\delta} \phi \right) {^\ast\!} R^{\delta ( \mu \nu ) \sigma} , \\
  T^{\phi}_{\mu \nu} &\equiv \left( \nabla_{\mu} \phi \right) \left( \nabla_{\nu} \phi \right) - \frac{1}{2} g_{\mu \nu} \nabla_{\delta} \phi \nabla^{\delta} \phi  - g_{\mu \nu} V(\phi),
\end{align}
where the indices enclosed by parentheses in superscripts are supposed to be symmetrized. 

As to the metric, we take a flat Friedmann--Lema{\^i}tre--Robertson--Walker model with GW perturbation, $h_{i j}$:
\begin{equation}
d s^2 = -d t^2 + a(t)^2\left(\delta_{i j}+h_{i j}\right)d x^i d x^j \ ,
\end{equation}
where $a(t)$ is the scale factor. Hereafter, we ignore the cosmic expansion until Sec.~\ref{Sec:Condition for the GW resonance}, where we discuss the evolution of axion in the history of the universe. Throughout this paper, we assume that GWs are propagating in the $x$--direction, for definiteness. Furthermore, we assume that the variation of the axion field in the $y$-- and $z$--directions can be neglected, {\it i.e.}, we neglect the diffraction of GWs. 

We employ the following ansatz for the axion field:
\begin{align}
  \label{Eq:axion}
  \phi(t, x) = \frac12 \left( \varphi(x) e^{- i m t} + \varphi^{\ast}(x) e^{i m t} \right),
\end{align} 
where $\varphi(x)$ denotes the spatial distribution of the amplitude of the axion oscillations including the phase and ``$\ast$'' represents the complex conjugation. Throughout this paper, we assume that in axion clouds, where the axion has a much larger amplitude $|\varphi(x)|$ than the average value in the present universe, the axion is virialized and its coherence length is roughly given by the de Broglie wavelength $\lambda_{\rm c} = 2 \pi/m v$. Then, the spatial derivative of $\phi$ is much smaller than its time derivative:
\begin{align}
  \label{Eq:vvsm}
\frac{|\phi'|}{|\dot \phi|} \sim \frac{1}{m \lambda_{\rm c}} = \frac{v}{2 \pi} \ll 1 \,,
\end{align}
where the dot and the prime denote the differentiations in $t$ and $x$, respectively, and $v \ll 1$ is the velocity dispersion of the axion, which can be approximated by the virial velocity of the corresponding cloud. If we assume that our galaxy consists of the axion, the fiducial value of $v$ would be
\begin{equation}
  v \approx 7 \times 10^{-4}.
\end{equation}
Therefore, Eq.~\eqref{Eq:axion} is the solution of the approximate evolution equation for the axion:
\begin{align}
\ddot{\phi} = - \frac{d V}{d \phi} 
\quad \text{with} \quad V = \frac{1}{2}m^2 \phi^2 \,,
\end{align}
where we neglect the Pontryagin density $\pont$ in the right--hand side. We assume that there is no source of generating a large amplitude of $\pont$ in the universe. Although we later discuss resonant amplification of GWs caused by axion oscillations, even in this case the net production of $\pont$ will be significantly suppressed since the direction of the circular polarization of the amplified GWs oscillates according to the phase of the axion oscillation. Furthermore, the energy density of gravitational waves decays faster than that of the axion cloud. Hence, the effect of the background gravitational waves on the dynamics of the axion cloud would be expected to be small in general. Besides the effect of $\pont$ contained in the GW background, When we consider the situation in which GWs are efficiently amplified in the axion cloud, the backreaction to the axion dynamics would be significant at some point. To discuss beyond that point, we would need to take into account the decay of axion oscillations by considering the energy balance, which simply decelerates the further amplification of GWs. Therefore, it would be reasonable to assume that $\pont$ can be neglected in most cases of interest. It is convenient to define the dimensionless field as
\begin{align}
\label{Eq:epsilondef}
  \varepsilon(x) \equiv \frac{\sqrt{2} \ell_{\rm dCS}^2}{\Mpl} m^2 \varphi(x) 
\end{align}
In terms of the local energy density of the axion cloud, 
\begin{align}
\rho_{\rm a} = T_{\phi}^{0 0} = \frac12 m^2 |\varphi|^2 
\end{align}
where we ignore the contribution from the spacial derivative terms, the dimensionless amplitude is expressed as 
\begin{align}
  \label{Eq:epsilon_order0}
 |\varepsilon| = 2 \ell_{\rm dCS}^2 \frac{m}{\Mpl} \sqrt{\rho_{\rm a}} \,. 
\end{align}
The equation of motion for GWs in a flat background in dCS--axion gravity can be written as
\begin{equation}
  \label{Eq:EoMofh}
    ( \partial_t^2 - \partial_x^2 ) h_{\rm R/L} - \frac12 i \lambda_{\rm R/L} \left( \varepsilon(x) e^{- i m t} + \varepsilon^{\ast}(x) e^{i m t} \right) \partial_t \partial_x  h_{\rm R/L} = 0,
\end{equation}
where the indices R and L represent the right--handed and the left--handed polarization modes, respectively. $\lambda_{\rm R} = + 1$ and $\lambda_{\rm L} = - 1$, and we neglect higher order dCS corrections and the spatial derivatives of $\phi$ because of Eq.~\eqref{Eq:vvsm}.

Next, we review a simple description of the parametric resonance in the case with $\varepsilon(x) = \rm const.$ Let us consider a plane wave propagating in the $x$--direction, {\it i.e.}, $h_{\rm R/L}(t, x) = A_0 e^{- i\omega t + i k x} + {\rm c.c.}$, where $A_0$ is constant, ``${\rm c.c.}$'' denotes the complex conjugate and $\omega$ and $k$ are the angular frequency and the angular wavenumber, respectively. The axion field oscillating at frequency $m$ mediates the interaction between this plane wave and another one whose frequency is lower by $m$, {\it i.e.}, $h_{\rm R/L}(t,x) = B_0 e^{- i ( \omega - m ) t + i k x} + {\rm c.c.}$ with a constant $B_0$, and causes the resonance in a narrow frequency band around $\omega\approx k\approx m/2$. Although the axion interaction also excites metric perturbations at higher frequencies with $\omega\approx (n+1/2)m$, where $n$ is a natural number, these higher frequency modes are not independently propagating modes and their amplitudes are suppressed. Therefore we ignore them, and thus, we start with the following ansatz:
\begin{align}
  \label{Eq:hansatz0}
  h_{\rm R/L}(t, x) = \left( A_0 \, e^{- i m t / 2} + B_0 \, e^{i m t / 2} \right) e^{- i \domega t + i ( \dk + m / 2 ) x} \,,
\end{align}
where $\domega = \omega - m / 2$ and $\dk = k - m / 2$ are introduced. The first term associated with $A_0$ represents a right--going wave, while the second term with $B_0$ propagates in the opposite direction. Then, the equation of motion for GWs~\eqref{Eq:EoMofh} reduces to a set of two algebraic equations:
\begin{align}
  \varepsilon B_0  = 8 i \lambda_{\rm R/L} \frac{\dk - \domega}{m}A_0 \,, \qquad
  \varepsilon^* A_0  = - 8 i \lambda_{\rm R/L} \frac{\dk + \domega}{m}B_0 \,,
\end{align}
where $\domega, \dk \ll \mathcal{O}(m)$ is assumed. Hence, the dispersion relation reads
\begin{align}
  \label{Eq:dispersionrelationori}
  \domega^2 = \dk^2 - \frac{m^2}{64} |\varepsilon|^2 \,.
\end{align}
If $\dk$ is in a narrow band
\begin{equation}
\label{Eq:resonantband}
    |\dk| < \frac{m |\varepsilon|}{8} \,,
\end{equation} 
$\domega^2$ becomes negative, which means the resonant instability occurs. 

In order to take account of the spatial distribution of the axion in the following sections, we generalize the ansatz~\eqref{Eq:hansatz0} as
\begin{align}
  \label{Eq:hansatz}
  h_{\rm R/L}(t, x) = \left( A(x) \, e^{- i m t / 2} + B(x) \, e^{i m t / 2} \right) e^{- i \domega (t - x) + i m x / 2} \,,
\end{align}
where we set $\dk = \domega$ since the deviation of $\dk$ from $\domega$ can be absorbed by $A(x)$ and $B(x)$. The equation of motion~\eqref{Eq:EoMofh} with Eq.~\eqref{Eq:hansatz} reduces to a set of two differential equations:
\begin{align}
  \label{Eq:approxEq1}
  &A' - \frac{m}{8} \lambda_{\rm R/L} \varepsilon B  = 0 \,, \\
  \label{Eq:approxEq2}
  &B' + 2 i \domega B + \frac{m}{8} \lambda_{\rm R/L} \varepsilon^* A= 0 \,.
\end{align}
By eliminating $B(x)$ from these equations and rewriting the resulting equation in terms of $X(x) \equiv A'/A$, we obtain
\begin{align}
  \label{Eq:masterEqX}
  X' + X^2 + \left( 2 i \domega - \frac{\varepsilon'}{\varepsilon} \right) X + \frac{m^2}{64} |\varepsilon|^2 = 0 \,.
\end{align}
We will solve the above equation in two different limits in Sec.~\ref{Sec:coherentcase} and \ref{Sec:the effect of spatial distribution}.

\section{Coherent Axion Field}
\label{Sec:coherentcase}

\subsection{Resonant instability of a wave packet}
\label{Sec:CoherentWvepackt}

First, let us consider the simple case in which the axion oscillates almost coherently. More precisely, we assume $|\varepsilon| \gg |( \ln \varepsilon )'/m| \sim v$, and hence we neglect the $\varepsilon'$--term in Eq.~\eqref{Eq:masterEqX}. Furthermore, we neglect the $X'$--term, which will turn out to be consistent immediately below. Then, the equation of motion to solve becomes
\begin{align}
  \label{Eq:resonanteq}
  X^2 + 2 i \domega X + \frac{1}{64} m^2 |\varepsilon|^2= 0 \,.
\end{align}
The solution is obtained as
\begin{align}
  \label{Eq:Xcohe}
  X(x) = - i \left( \domega - \sqrt{\domega^2 + \frac{m^2 |\varepsilon|^2}{64}} \right) \,,
\end{align}
where the branch cut runs from $\domega = - i m |\varepsilon| / 8$ to $\domega = + i m |\varepsilon| / 8$. The sign of the square root is chosen so that $X(x)$ becomes zero in the limit $|\domega| \gg m |\varepsilon| / 8$. Now, we can confirm
\begin{align}
\left| \frac{[\ln (X/m)]'}{m} \right| = \left| - \frac{m}{128} \frac{\left( \varepsilon^* \varepsilon' + \varepsilon^*{}' \varepsilon \right)}{\left( \domega - \sqrt{\domega^2 + m^2 |\varepsilon|^2 / 64} \right) \sqrt{\domega^2 + m^2 |\varepsilon|^2 / 64}} \right| \sim v \ll |\varepsilon| \,,
\end{align}
which justifies neglecting $X'$ in Eq.~\eqref{Eq:resonanteq}. Thus, the right--going plane wave is proportional to
\begin{equation}
\label{eq:hbranchcut}
  h_{\domega} \propto \exp \left[ - i \domega \left( t - x \sqrt{1 + \frac{|\varepsilon|^2}{64} \frac{m^2}{\domega^2}} \right) \right] \,,
\end{equation}
where we have abbreviated the common phase factor $e^{- i m ( t - x ) / 2}$ and the contribution from the $B(x)$--term. As mentioned above, the fastest growth of the amplitude is achieved at $\domega = i m |\varepsilon| / 8$. In this case, therefore, the the metric perturbation behaves as
\begin{equation}
  \label{eq:resonantGrowth}
  h_{\domega} \propto \exp( - i \domega t ) = \exp \left[ \frac{m |\varepsilon|}{8} t \right] \,.
\end{equation}

So far, we have discussed the resonant amplification of a plane GW. However, the analysis on a plane wave is not enough to understand how the gravitational waveform passing through an axion cloud is affected by the resonance. In order to address this issue, let us consider a wave packet propagating in an axion cloud which has a vanishing amplitude $\varepsilon = 0$ for $x < 0$ and coherently oscillates with a finite amplitude, {\it i.e.}, $\varepsilon' = 0$, for $x > 0$. Though we assume the axion amplitude $|\varepsilon(x)|$ smoothly changes around $x = 0$, we focus on the asymptotic region with sufficiently large $x$ and $t$, which will allow us to avoid the complexity associated with the transition around $x = 0$. In the asymptotic region where the above analysis on a plane wave applies, each Fourier mode of $h$ is approximately given by Eq.~\eqref{eq:hbranchcut}. We consider a Gaussian wave packet given by a superposition of waves at different frequencies with the weight $\propto \exp(-\delta\omega^2/2K^2)$, where the constant $K^{-1}$ represents the width of the wave packet. Integrating over the frequency, we find that a simple Gaussian wave packet in the region of $x<0$ ends up with 
\begin{equation}
  \label{Eq:wavepacket}
  h_{\rm packet} = \frac{1}{\sqrt{2 \pi} K} e^{- i m ( t - x ) / 2} \, \int \dd \domega \, e^{f(\domega)} \,,
\end{equation}
where
\begin{align}
  \label{Eq:phaseofpacket}
  f(\domega) = - i \domega \left( t - x \sqrt{1 + \frac{|\varepsilon|^2}{64} \frac{m^2}{\domega^2}} \right) - \frac{\domega^2}{2 K^2} \,,
\end{align}
in the region of $x>0$. 

Evaluating the integral in Eq.~\eqref{Eq:wavepacket} with the method of the steepest decent, we obtain the leading--order expression for $h_{\rm packet}$ as 
\begin{subnumcases}
{h_{\rm packet}\sim}
\label{Eq:pachetfuture}
\exp \left[ \dfrac{m |\varepsilon|}{8} \sqrt{t^2 - x^2} \right]\,,\qquad (t > x) \,, & \\
\label{Eq:pachettip}
\exp \left[ \dfrac{3}{32} \left( \dfrac{m^2 |\varepsilon|^2 t}{2 K} \right)^{2/3} \right] \,, \qquad (x = t)\,, & \\
\label{Eq:packetacausal}
\exp \left[ - \dfrac{1}{2} K^2 (x - t)^2 \right] \,, \qquad (t < x) \,, &
\end{subnumcases}
where the common exponential factor $e^{- i m ( t - x ) / 2}$ and prefactors are suppressed. The derivation and the complete expressions can be found in Appendix~\ref{sec:Calculation of the wave packet}. One can observe in Eq.~\eqref{Eq:pachetfuture} that the amplification of the GWs is dominant in the causal future after a long time. From Eq.~\eqref{Eq:packetacausal}, one can immediately confirm that the resonant amplification is absent in the acausal region $t < x$, as expected. Although Eq.~\eqref{Eq:pachetfuture} may appear to imply that arbitrary small GWs can be exponentially amplified, it is not the case when the spatial extension of the axion cloud is small enough, which will become obvious from the discussion in the next subsection.

\begin{figure}[tbp]
  \begin{minipage}{0.45\hsize}
    \begin{center}
      \includegraphics[width=70mm]{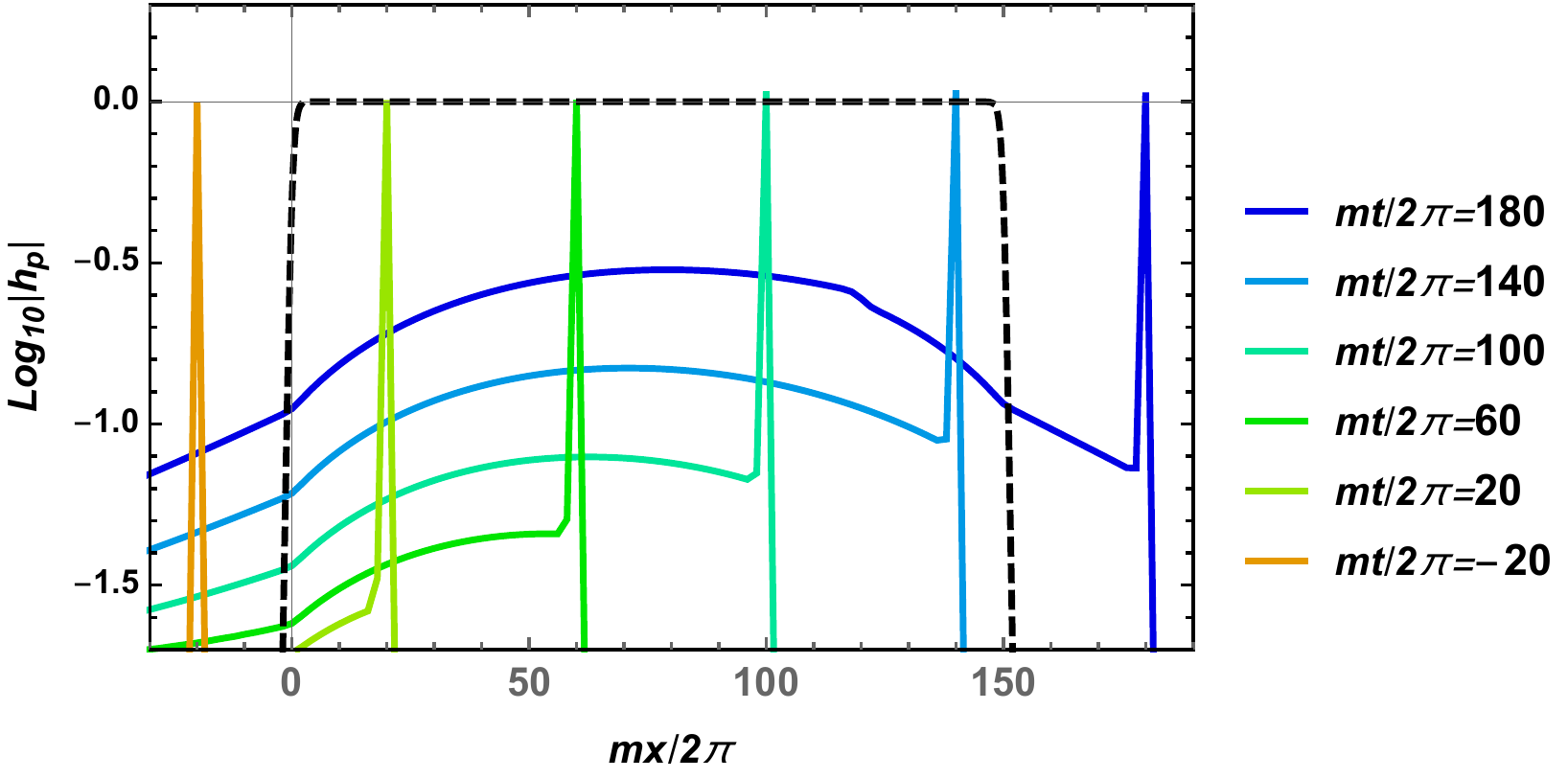}
    \end{center}
  \end{minipage}
  \begin{minipage}{0.45\hsize}
    \begin{center}
      \includegraphics[width=70mm]{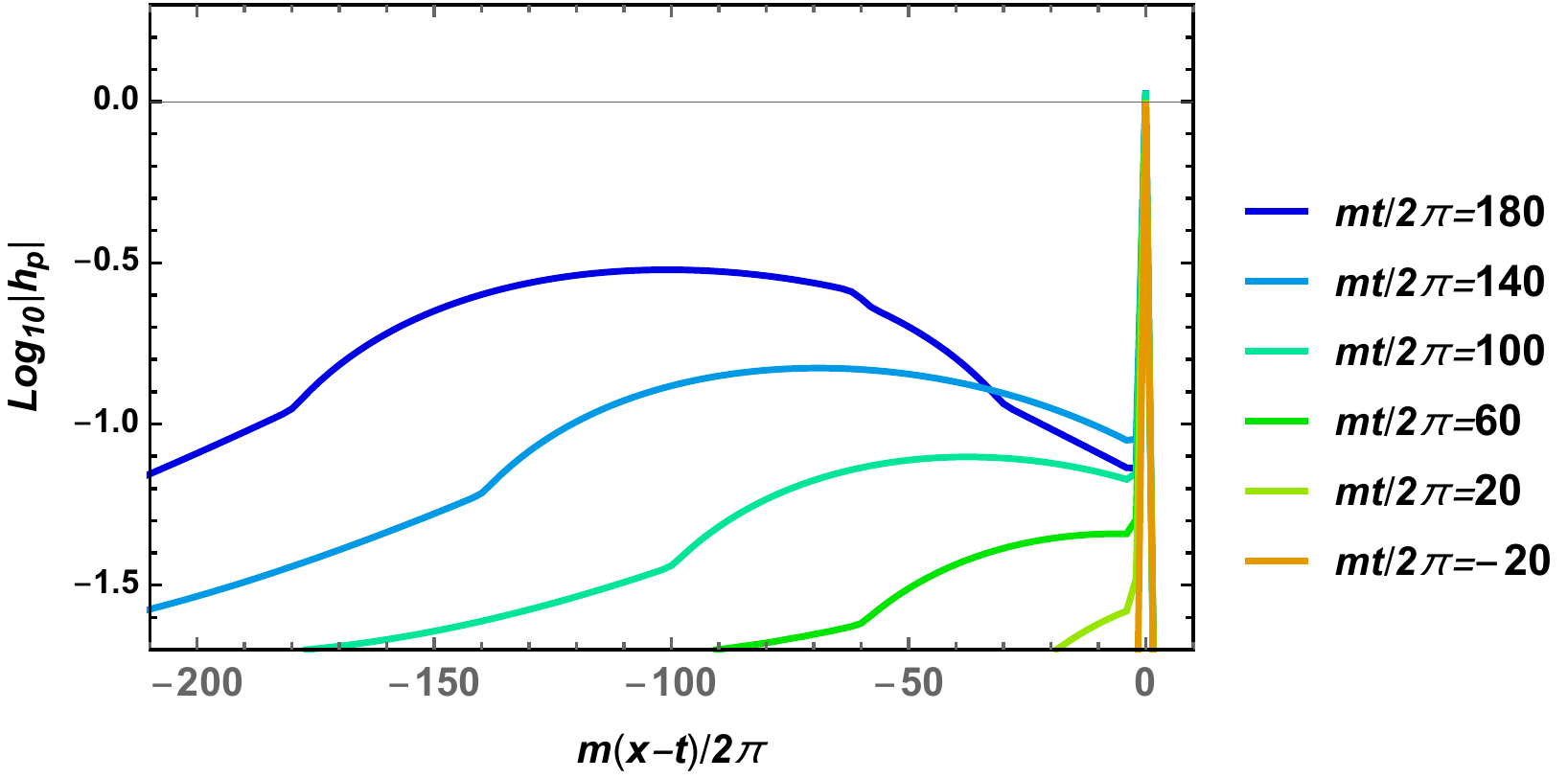}
    \end{center}
  \end{minipage}
  \caption{The time evolution of the wave packet. The left panel shows how the wave packet propagates in the axion cloud, while the horizontal axes is $m(x - t)/2\pi$ in the right panel such that the original center of the wave packet is overlapped. The black dashed line in the left panel denotes the distribution of the axion amplitude $|\varepsilon(x)|$ whose plateau value is $0.03$. The resonant amplification is significant in the coherent axion cloud but it takes place only in the causal future and slightly on the tip of the wave packet, as we have analytically shown in Eqs.~\eqref{Eq:pachetfuture}-\eqref{Eq:packetacausal}.}
  \label{Fig:resonantGW1}
\end{figure}

\begin{figure}[tbp]
  \begin{minipage}{0.45\hsize}
    \begin{center}
      \includegraphics[width=70mm]{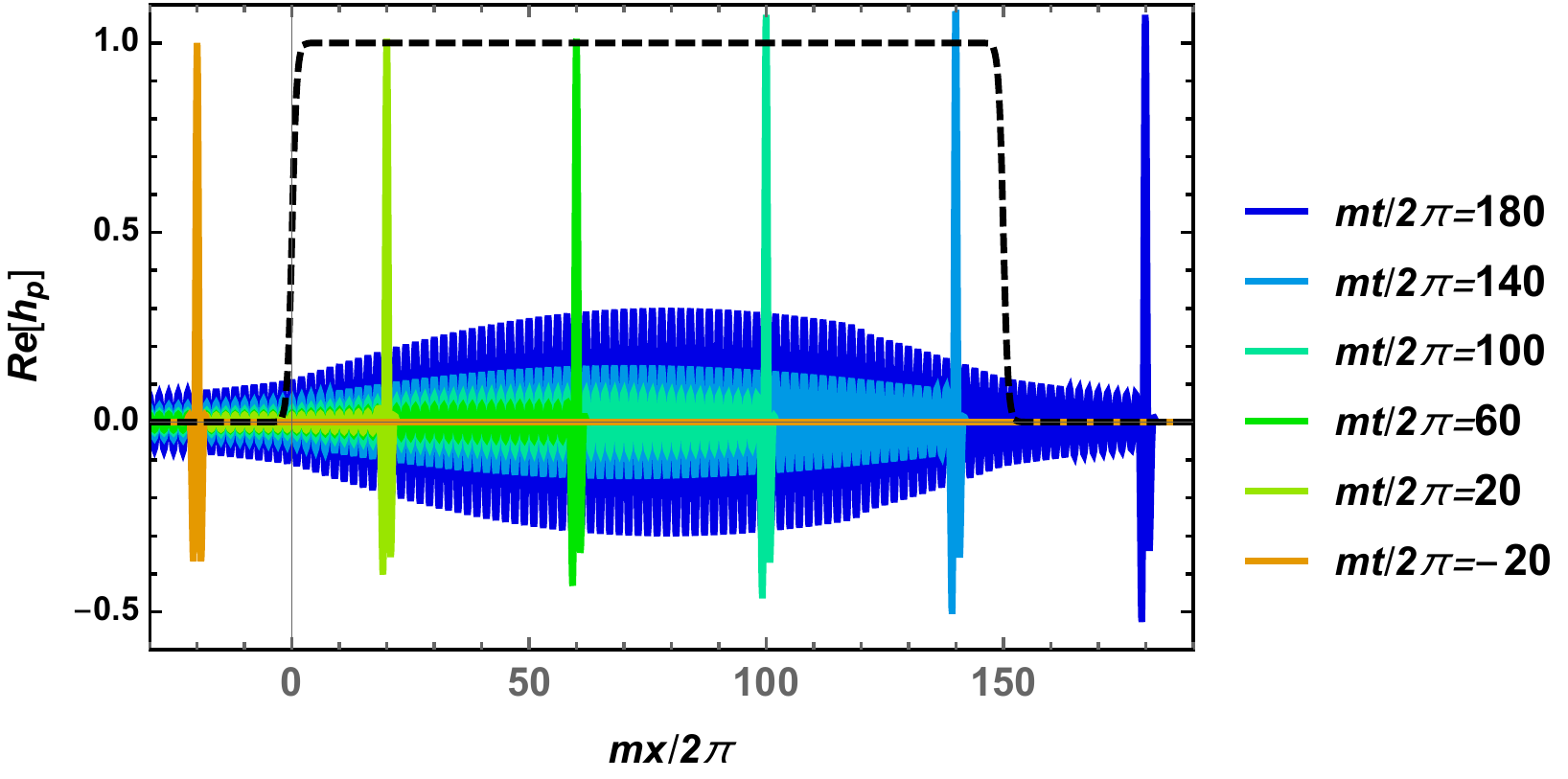}
    \end{center}
  \end{minipage}
  \begin{minipage}{0.45\hsize}
    \begin{center}
      \includegraphics[width=70mm]{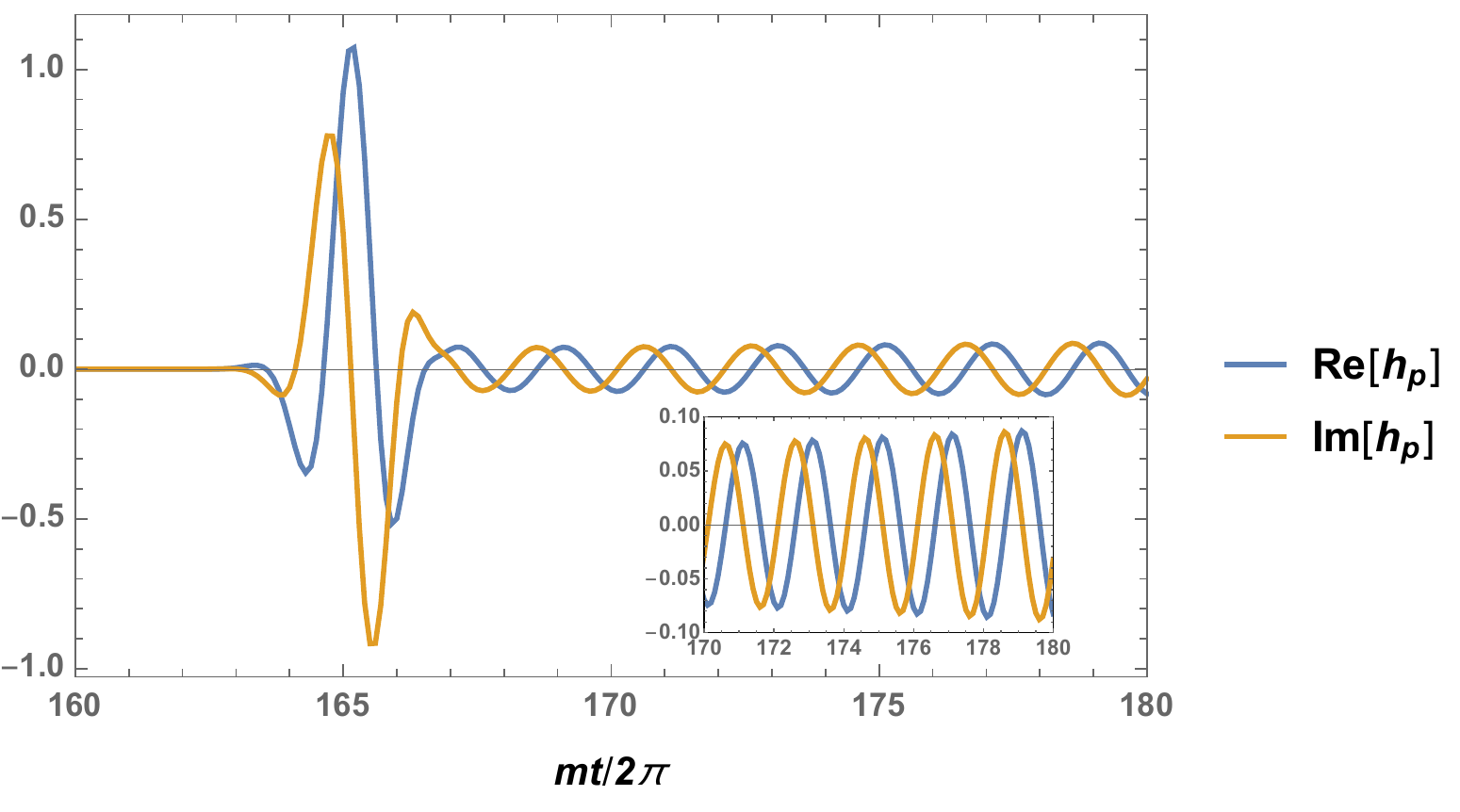}
    \end{center}
  \end{minipage}
  \caption{The left panel shows the real part of the same wave packet as Fig.~\ref{Fig:resonantGW1}. The black dashed line in the left panel shows the axion cloud distribution $|\varepsilon(x)|$. One can see in the left panel that the coherent axion continues to amplify GWs even after the incident wave packet leaves the cloud. The right panel shows how the GWs would be observed as the function of time outside the axion cloud at $x = 165 \times 2 \pi m^{- 1}$. One can find that the amplitudes of subsequent waves gradually increase due to the continuous amplification in the cloud.}
  \label{Fig:realGW1}
\end{figure}

We numerically calculate the evolution of the wave packet directly solving Eq.~\eqref{Eq:EoMofh} to compare with the above analytic results. We set the initial condition for the normalized wave packet as $h_{\rm R} = \exp[- i m ( t - x ) / 2 - K^2 ( t - x )^2 / 2]$ with  $K = m / 4$ and solve its evolution from  $t = - 20 \times 2 \pi m^{- 1}$ to $t = 180 \times 2 \pi m^{- 1}$. The coherent axion cloud extends from $x \approx 0$ to $x \approx 150 \times 2 \pi m^{- 1}$, its amplitude is $|\varepsilon| = 0.03$, and the inside and the outside of the cloud are smoothly connected by the sigmoid function, $|\varepsilon| \propto [\tanh(x) \tanh(x_{\rm end} - x) + 1 ] / 2$, with the right--boundary of the axion cloud $x = x_{\rm end}$, beyond which $\varepsilon$ decays. Figures~\ref{Fig:resonantGW1} and \ref{Fig:realGW1} show the time evolution of the wave packet. Note that we employed the value of the axion amplitude much larger than the upper bound that we will obtain in Eq.~\eqref{Eq.marginalcond} to demonstrate the amplification effect. In these figures, one can also observe that the resonant amplification of GWs by a coherent axion cloud is significant in the causal future but not in the acausal region.

It is worth mentioning that our results on the resonant GWs in the coherent case correct a little misleading statement on the arrival time--delay of the resonance in Ref.~\cite{Jung:2020aem} that the arrival of GWs in the resonant frequency band delays relative to that of other frequencies. What the resonance effect causes is not a simple frequency--dependent time--delay. Therefore, the constraints derived based on this picture are required to be re-examined.

\subsection{Critical amplitude of coherent axion cloud}

Here, we discuss the condition for GWs to grow inside an axion cloud without any GW inflow from the outside. For definiteness, we assume a coherent axion cloud whose amplitude $|\varphi|$ is constant, {\it i.e.}, $\varepsilon' = 0$, over the region $0<x<L$ and rapidly decays outside this region. This setup might be too idealized to describe a realistic cloud, but it can give a good approximation if we focus on each coherent patch as well as the early universe when the axion starts to oscillate and the coherence length can be sufficiently large. We will obtain the threshold of the axion's amplitude for the resonant instability of GWs to occur.

To find the condition for a growing mode to exist, we set $\Re[\domega]=0$ and consider non--vanishing imaginary part of $\domega$, {\it i.e.}, $\omega_I:=\Im[\domega] > 0$. Then, the solution for $A(x)$ inside the axion cloud is obtained from Eq.~\eqref{Eq:Xcohe} as
\begin{equation}
  \label{Eq:SolAofx}
  A(x) = A_+ \exp\left[\frac{m|\varepsilon|}{8} \left(\Omega+\sqrt{\Omega^2-1}\right)x \right] +A_- \exp\left[\frac{m|\varepsilon|}{8} \left(\Omega - \sqrt{\Omega^2-1}\right)x \right]\,,
\end{equation}
where $A_\pm$ are integration constants and we introduced $\Omega \equiv 8 \omega_I / (m |\varepsilon|)$. We impose the purely outgoing boundary conditions at both boundaries, {\it i.e.},
\begin{equation}
  A(0) = 0, \qquad B(L) = 0 .
\end{equation}
The first condition requires $A_- = - A_+$. Since $B(x) \propto A'(x)$ follows from Eq.~\eqref{Eq:approxEq1}, the second condition is recast into $A'(L)=0$, which reads
\begin{equation}
  \label{Eq:conddelta}
  \tan \left( \frac{m |\varepsilon|}{8} L \sqrt{1 - \Omega^2}  \right) = - \frac{\sqrt{1 - \Omega^2}}{\Omega} \,.
\end{equation}
The trivial solution of the above equation is $\Omega = 1$. In this case, however, we merely obtain $A(x)=B(x)=0$, so that no GW is induced. A non--trivial solution can be found for $0<\Omega<1$.%
\footnote{For $\Omega>1$, Eq.~\eqref{Eq:conddelta} can be rewritten as $\tanh(m |\varepsilon| L \sqrt{\Omega^2-1}/8)=-\sqrt{\Omega^2-1}/\Omega$. In this equation, the left hand side is positive, while the right hand side is negative. Thus, no solution is found for $\Omega>1$.}
Since the right hand side of Eq.~\eqref{Eq:conddelta} is negative for $0<\Omega<1$, the argument of the tangent function has to take a value between $(1/2+n)\pi$ and $(1+n)\pi$ where $n=0,1,2,\cdots$. Then the lowest axion amplitude for the resonant amplification is achieved when $m|\varepsilon|L/8$ is equal to $\pi/2$. When $m|\varepsilon|L/8$ is slightly larger than $\pi/2$, as $\Omega$ increases from zero to unity, the left hand side in Eq.~\eqref{Eq:conddelta} decreases from a finite value towards negative infinity, while the right hand side increases from negative infinity to zero. Thus, there must be a solution for Eq.~\eqref{Eq:conddelta}. As a result, we obtain the condition for GWs to exponentially grow inside the axion cloud with the size $L$ as
\begin{equation}
  \label{Eq:condspontaneous}
  |\varepsilon|>\varepsilon_{\rm cri} \equiv \frac{4 \pi}{m L} \,,
\end{equation}
where $\varepsilon_{\rm cri}$ denotes the threshold amplitude of the axion cloud. If this condition is satisfied, there exists a GW solution in the following form:
\begin{equation}
    h_{\rm R/L}(t,x) = 2i A_+ e^{-i\frac{m}{2}(t - x)}
    \exp\left(\frac{m|\varepsilon|}{8}\Omega\, t\right)
    \sin\left(\frac{m|\varepsilon|}{8}\sqrt{1-\Omega^2}\, x \right)
    + \cdots\,,
\end{equation}
where $\Omega$ in the range $0<\Omega<1$ is specified as a solution of Eq.~\eqref{Eq:conddelta}, and we suppressed the contribution from the left--going component $B(x)=8 A'(x)/(m\varepsilon\lambda_{\rm R/L})$, which is the same order of magnitude as the right--going component. Although the above expression is valid only inside the axion cloud, $0<x<L$, it does not vanish at $x=L$ which infers the axion cloud spontaneously emits GWs and the resonant instability of the axion cloud occurs. Of course, the axion cloud remains stable if the dCS coupling is turned off since the instability is caused by the energy transfer from the axion to GWs.

\section{Incoherent Axion Field}
\label{Sec:the effect of spatial distribution}

Since axion clouds are expected to be virialized in high density regions and has only a finite coherence length in the present universe as discussed around Eq.~\eqref{Eq:vvsm}, it is essential to relax the assumption of its perfect coherence made in the previous section to study more realistic GW amplification. In this section, therefore, we take into account the spatial distribution of the axion field, {\it i.e.}, $\varepsilon'(x) \neq 0$. Since it later turns out that a consistent solution can be found by neglecting $X^2$--term, we solve the simplified equation
\begin{align}
  \label{Eq:approxEqX2}
   X' + \left( 2 i \domega - \frac{\varepsilon'}{\varepsilon} \right) X + \frac{m^2}{64} |\varepsilon|^2 = 0 \,.
\end{align}
The solution is given by
\begin{equation}
    X(x) = e^{- 2 i \domega x} \varepsilon(x) \left[ C_0 + \frac{m^2}{64} \int_x^{x_{\rm end}} \dd y \, \varepsilon^*(y) e^{2 i \domega y} \right],
\end{equation}
where $C_0$ is the integration constant and $x_{\rm end}$ is the right--boundary of the axion cloud beyond which $\varepsilon(x)$ vanishes. The integration constant can be fixed by requiring the boundary condition $B(x_{\rm end}) = 0$, which means that there is no left--going wave from the past infinity. From Eq.~\eqref{Eq:approxEq1}, one finds
\begin{align}
  \left. \frac{B}{A} \right|_{x = x_{\rm end}} = \left. \frac{8 \lambda_{\rm R/L} X}{m \varepsilon} \right|_{x = x_{\rm end}} = \frac{8 \lambda_{\rm R/L}}{m} C_0 e^{- 2 i \domega x_{\rm end}} = 0 \,.
\end{align}
Then, the integration constant $C_0$ is fixed to 0. Now, we find that the solution at the leading order is given by
\begin{align}
  \label{Eq:solX}
  X(x) &= \frac{m^2}{64}  \int_x^{x_{\rm end}} \dd y\, \varepsilon(x) \varepsilon^*(y) e^{- 2 i \domega ( x - y)}  \,.
\end{align}
Immediately, substituting Eq.~\eqref{Eq:solX} into Eq.~\eqref{Eq:approxEq1}, we obtain 
\begin{align}
  \label{Eq:incoherentB}
  B(x) = \frac{m}{8} \lambda_{\rm R/L}  \int_x^{x_{\rm end}} \dd y \, \varepsilon^*(y) e^{- 2 i \domega ( x - y )} \,.
\end{align}
Therefore, a left--going wave of $\mathcal{O}(\varepsilon)$ is produced. 

In order to evaluate the integral in Eq.~\eqref{Eq:solX} and to observe the evolution of the wave packet, we assume a Gaussian distribution of the axion whose coherence length is $\lambda_{\rm c}$, 
\begin{align}
\label{Eq:gaussepsilon}
  \left\langle \varepsilon(x) \varepsilon^*(y) \right\rangle 
  = |\bar \varepsilon|^2 \exp \left[ - \frac{( x - y )^2}{2 \lambda_{\rm c}^2} \right] \,,
  \quad
  \left( x, y \in [x_0, x_{\rm end}] \right) \,, 
\end{align}
where $\langle \cdots \rangle$ is the ensemble average. We set the left--boundary of the axion cloud at $x = x_0$ $(< x_{\rm end})$. Also, assuming $\langle |\varepsilon(x)|^2 \rangle$ in the cloud is constant, we denote it by $|\bar \varepsilon|^2$. From the definition $X\equiv A'/A$, $\ln A$  is given by the $x$ integral of $X(x)$. The ensemble average of $\ln A$ is computed as
\begin{align}
\label{Eq:MeanX}
\left\langle \ln A(x) \right\rangle 
&= \left\langle \int^x_{x_0} \dd y \, X(y) \right\rangle \,, \notag\\
&= \frac{m^2}{64} |\bar \varepsilon|^2 \int^x_{x_0} \dd y \, \int^{x_{\rm end} - y}_0 \dd z \, e^{- \frac{z^2}{2 \lambda_{\rm c}^2} + 2 i \domega z} \,, \notag\\
&= \frac{\sqrt{\pi/2}}{64} m^2 \lambda_{\rm c} e^{- \hdomega^2} |\bar \varepsilon|^2 \int^x_{x_0} \dd y \, \left[ \erf \left( \frac{x_{\rm end} - y}{\sqrt{2} \lambda_{\rm c}} - i \hdomega \right) + \erf \left( i \hdomega \right) \right] \,, \notag\\
&\simeq \frac{\sqrt{\pi/2}}{64} m^2 \lambda_{\rm c} e^{- \hdomega^2} \left[ 1 + \erf \left( i \hdomega \right) \right] |\bar \varepsilon|^2 (x - x_0) \,,
\end{align}
where $\hdomega \equiv \sqrt{2} \lambda_{\rm c} \domega$, and $x_{\rm end} - x \gg \lambda_{\rm c}$ is used in the last line. Here, the initial amplitude $A(x_0)$ is normalized to be unity. Since $|\domega|\lesssim m|\varepsilon|/8$ for an efficient resonance and $\lambda_{\rm c} = (m v/2\pi)^{-1}$, the magnitude of $\hdomega$ is tiny for $|\varepsilon|\ll v$. For later convenience, we define $\langle \ln A(x)\rangle$ in the limit $\hdomega \to 0$ as%
\footnote{In Ref.~\cite{Jung:2020aem}, since the authors discussed the incoherent case without solving the equations, they proposed two possibilities; the ``{\it maximum total enhancement}'' [Eq.~(23)] and the ``{\it minimum total enhancement}'' [Eq.~(24)]. Eq.~\eqref{Eq:lnAestimate} obtained by solving the equation of motion agrees with the former.}
\begin{align}
\label{Eq:lnAestimate}
\mathcal{C} &\equiv \left\langle \ln A(x) \right\rangle \big|_{\hdomega \to 0} \simeq \frac{\sqrt{\pi^3/2}}{32} \frac{m L |\bar \varepsilon|^2}{v} = \frac{\sqrt{\pi^3/2}}{8} \frac{\ell_{\rm dCS}^4 \, m^3 \, L \, \rho_{\rm a}}{v \, \Mpl^2} 
\quad \text{for} \quad x \gtrsim x_{\rm end} \,,
\end{align}
where $L\equiv x_{\rm end} - x_0$ is the size of the axion cloud, and  $e^{- \hdomega^2}[ 1 + \erf ( i \hdomega ) ] \xrightarrow{\hdomega \to 0} 1$ is used. Therefore, the resonant amplification is suppressed by $\mathcal{O}(\varepsilon/v)$ compared with the coherent case. However, we can still expect that GWs are significantly amplified for sufficiently large $\varepsilon$ and $L$. 

As we have 
\begin{align}
\langle \varepsilon(x) \varepsilon^*(y) \varepsilon(x) \varepsilon^*(z) \rangle &= 2 \langle \varepsilon(x) \varepsilon^*(y) \rangle \langle \varepsilon(x) \varepsilon^*(z) \rangle \,,
\end{align}
for Gaussian distribution, the ensemble average of $X^2(x)$ is evaluated as 
\begin{align}
\langle X^2(x) \rangle &\simeq \frac{m^2 |\bar{\varepsilon}|^2}{64} \times \frac{m^2 |\bar{\varepsilon}|^2 \lambda_{\rm c}^2}{64} \pi e^{- 2 \hdomega^2} \left[ 1 + \erf \left( i \hdomega \right) \right]^2 \,,
\end{align}
where we have assumed $x_{\rm end} - x \gg \lambda_{\rm c}$. Therefore, one can find that $\langle X^2(x) \rangle$ is also small enough compared with the source term in Eq.~\eqref{Eq:masterEqX}, {\it i.e.}, $m^2 |\varepsilon|^2 / 64$, as long as 
\begin{align}
m |\bar{\varepsilon}| \lambda_{\rm c} \ll 1 \,,
\end{align}
which is consistent with no spontaneous emission of GWs. Moreover, we stress that the expression for the amplification~\eqref{Eq:MeanX} obtained by neglecting the $X^2$--term is valid even if $\mathcal{C} > 1$, which is discussed in Appendix~\ref{Sec:AveragedSolution} in more detail.

We also calculate the variance of $\ln A$ as  
\begin{align}
\sigma_{\ln A}^2 
&\equiv \left\langle \left| \ln A(x) \right|^2 \right\rangle - \left| \left\langle \ln A(x) \right\rangle \right|^2  \notag\\
&= \frac{m^4}{4096} \int^x_{x_0} \dd y \int_y^{x_{\rm end}} \dd \tilde y \int_{x_0}^{x} \dd z \int_z^{x_{\rm end}} \dd \tilde z \, \langle \varepsilon(y) \varepsilon^*(\tilde z) \rangle \langle \varepsilon(z) \varepsilon^*(\tilde y) \rangle \, e^{- 2 i \domega ( y - \tilde y + z - \tilde z )} \notag\\
&\simeq \frac{\pi}{8192} m^4 \lambda_{\rm c}^2 e^{- 4 \lambda_{\rm c}^2 \domega^2} |\bar \varepsilon|^4 \int_{x_0}^{x} \dd y  \int_{x_0}^{x} \dd z \, \mathcal{G}(|y - z|) \,,\end{align}
with
\begin{equation}
    \mathcal{G}(|y - z|)\equiv
    \left[1-\erf \left(\frac{y - z}{\sqrt{2}\lambda_{\rm c}} + i \hdomega\right)\right]
    \left[1-\erf \left(\frac{z - y}{\sqrt{2}\lambda_{\rm c}} + i \hdomega\right)\right]\,,
\end{equation}
where $x_{\rm end} - x \gg \lambda_{\rm c}$ was assumed again. $\mathcal{G}(|y - z|)$ effectively constrains the integral range of $z$ into $z=[y - \sqrt{2}\lambda_{\rm c},y+\sqrt{2}\lambda_{\rm c}]$, because of its asymptotic behaviors
\begin{align}
    \mathcal{G}(|y - z|\ll \sqrt{2}\lambda_{\rm c})&\simeq \left[1 - \erf (i \hdomega)\right]^2,\\
    \mathcal{G}(|y - z|\gg \sqrt{2}\lambda_{\rm c})&\simeq \sqrt{\frac{8}{\pi}}\frac{\lambda_{\rm c}}{|y - z|}
    e^{-\left(\frac{|y - z|}{\sqrt{2}\lambda_{\rm c}} - i \hdomega\right)^2}\,.
\end{align}
Thus, the square root of variance ({\it i.e.}, standard deviation) in the limit $\hdomega \to 0$ is approximately estimated by using $\int_{-\infty}^{\infty} \dd x (1+\erf (x/\sqrt{2}\lambda_{\rm c}))(1+\erf (-x/\sqrt{2}\lambda_{\rm c}))= 4 \lambda_{\rm c}/\sqrt{\pi}$ as 
\begin{align}
\label{Eq:standarddeviation}
\sigma_{\ln A}
&\sim \, \frac{\pi^{7/4}}{4} \frac{\ell_{\rm dCS}^4}{\Mpl^2} \sqrt{\frac{m^5 \, L}{v^3}} \rho_{\rm a} \,. 
\end{align}
From the expression for the ratio $\sigma_{\ln A}/{\cal C} \simeq \pi^{-1/4}\sqrt{\lambda_{\rm c}/L}$, we find that the standard deviation is negligible compared to the mean value, since we assume that the cloud size $L$ is much larger than the coherence length $\lambda_{\rm c}$. Since the variance of $\ln A(x)$ can be neglected, the mean of $A(x)$ can be computed as 
\begin{align}
\label{Eq:AexpC}
\langle A(x) \rangle \simeq \exp \left( \mathcal{C} \right) \quad \text{for} \quad x \gtrsim x_{\rm end} \,.
\end{align}

\begin{figure}[tbp]
  \begin{minipage}{0.45\hsize}
    \begin{center}
      \includegraphics[width=70mm]{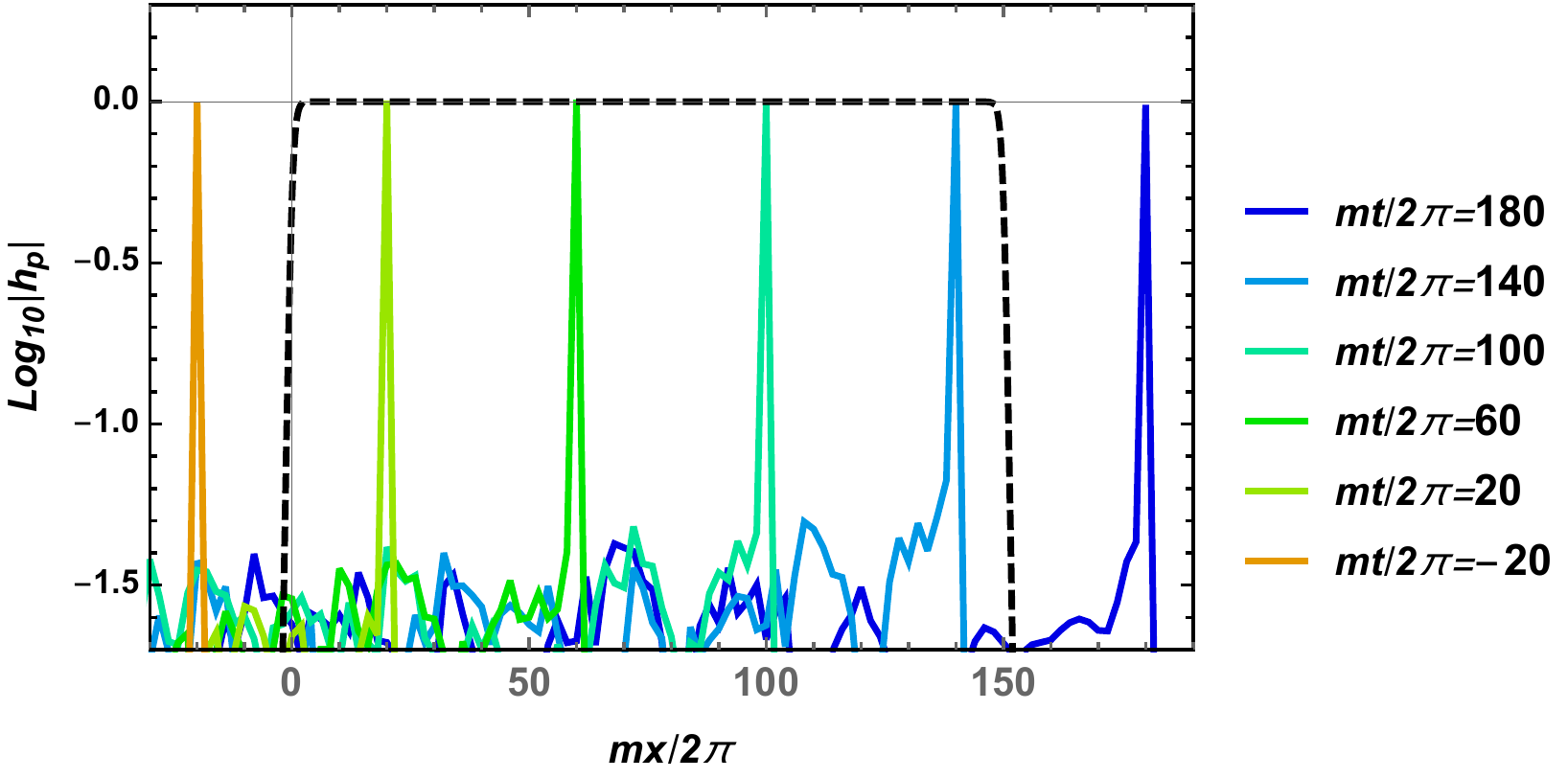}
    \end{center}
  \end{minipage}
  \begin{minipage}{0.45\hsize}
    \begin{center}
      \includegraphics[width=70mm]{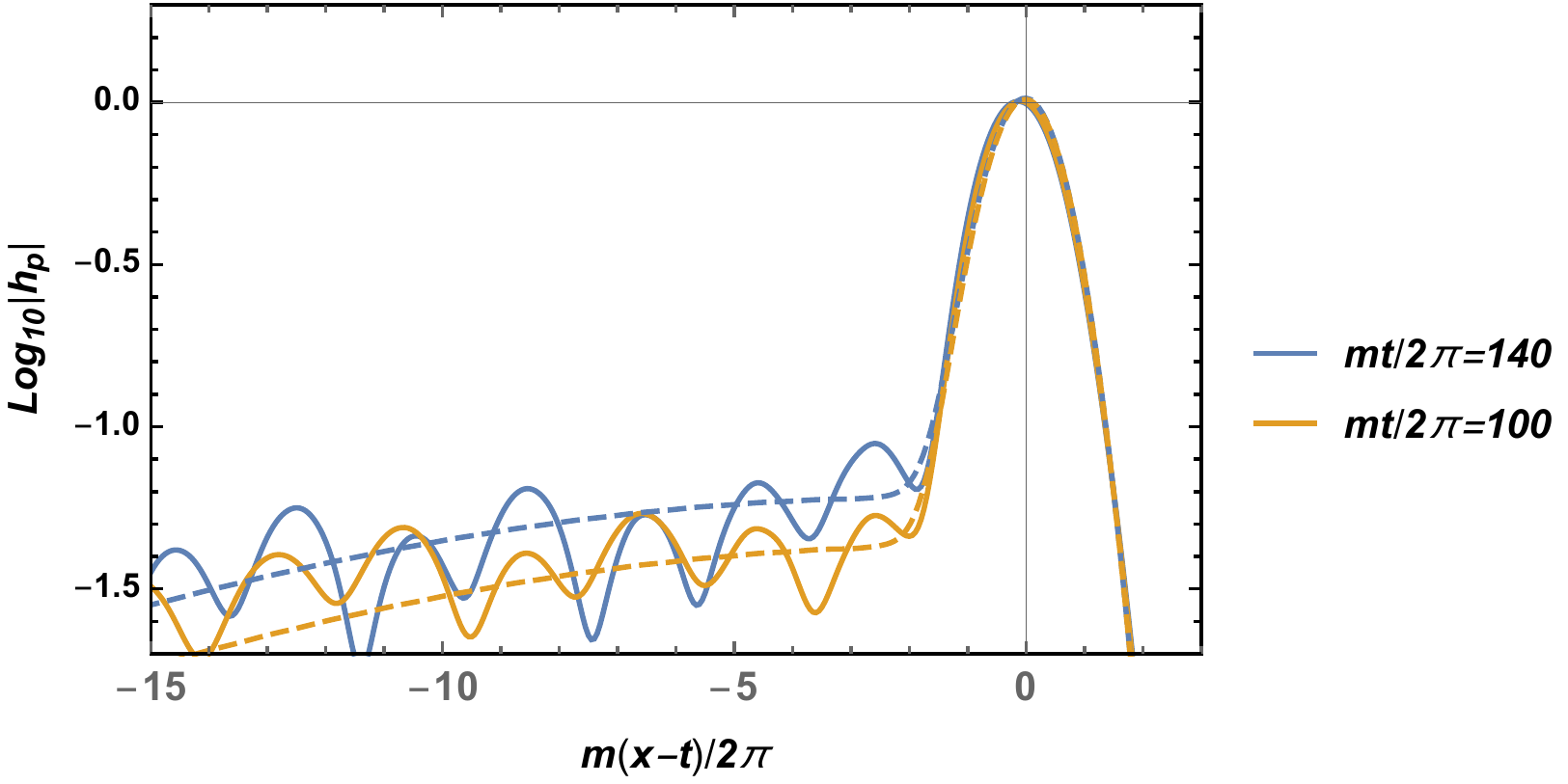}
    \end{center}
  \end{minipage}
  \caption{The time evolution of the wave packet. The left panel shows how the wave packet propagates in the axion cloud. The black dashed line in the left panel denotes the distribution of the axion amplitude $|\varepsilon(x)|$ whose plateau value is $0.03$. In the right panel, the horizontal axes is $m(x - t)/2\pi$, and hence the center of the wave packet is does not move. The dashed lines in the right panel denote the mean value of the wave packet over the axion phase realizations given by the numerical integration of Eq.~\eqref{Eq:hpincoh}. In contrast to the coherent case, the amplification in the causal future is less prominent.}
  \label{Fig:resonantGW2}
\end{figure}

The amplitude of the wave packet propagating in an incoherently oscillating axion cloud is expected to be given by 
\begin{align}
  \langle h_{\rm packet} \rangle
  &\propto e^{- \frac12 i m ( t - x )} \, \int \dd \domega \, e^{- \frac{\domega^2}{2 K^2} - i \domega ( t - x )} \langle A(x) \rangle \,, 
  \notag\\
  &\simeq e^{- \frac12 i m ( t - x )} \, \int \dd \domega \, e^{- \frac{\domega^2}{2 K^2} - i \domega ( t - x ) + \mathcal{C} e^{- \hdomega^2} \left( 1 + \erf (i \hdomega ) \right)} \,, 
  \notag\\
  &= \frac{1}{\sqrt{2} \lambda_{\rm c}} e^{- \frac12 i m ( t - x ) } \, \int \dd \hdomega \, e^{- \frac{\hdomega^2}{4 K^2 \lambda_{\rm c}^2}} \, e^{- i \frac{t - x}{\sqrt{2} \lambda_{\rm c}} \hdomega} \, e^{\mathcal{C} e^{- \hdomega^2} \left( 1 + \erf ( i \hdomega ) \right) } \,.
   \label{Eq:hpincoh}
\end{align}
To confirm the validity of this formula, the integral over $\hdomega$ is performed numerically and the result is compared with the direct numerical solution for Eq.~\eqref{Eq:EoMofh} in the right panel in Fig.~\ref{Fig:resonantGW2}. The dashed lines in the figure show the mean values Eq.~\eqref{Eq:hpincoh}, while the solid lines show the results in a realization of the axion's random distribution. One can observe that the latter moderately fluctuates around the former as expected in the figure.

\begin{figure}[tbp]
  \begin{minipage}{0.45\hsize}
    \begin{center}
      \includegraphics[width=70mm]{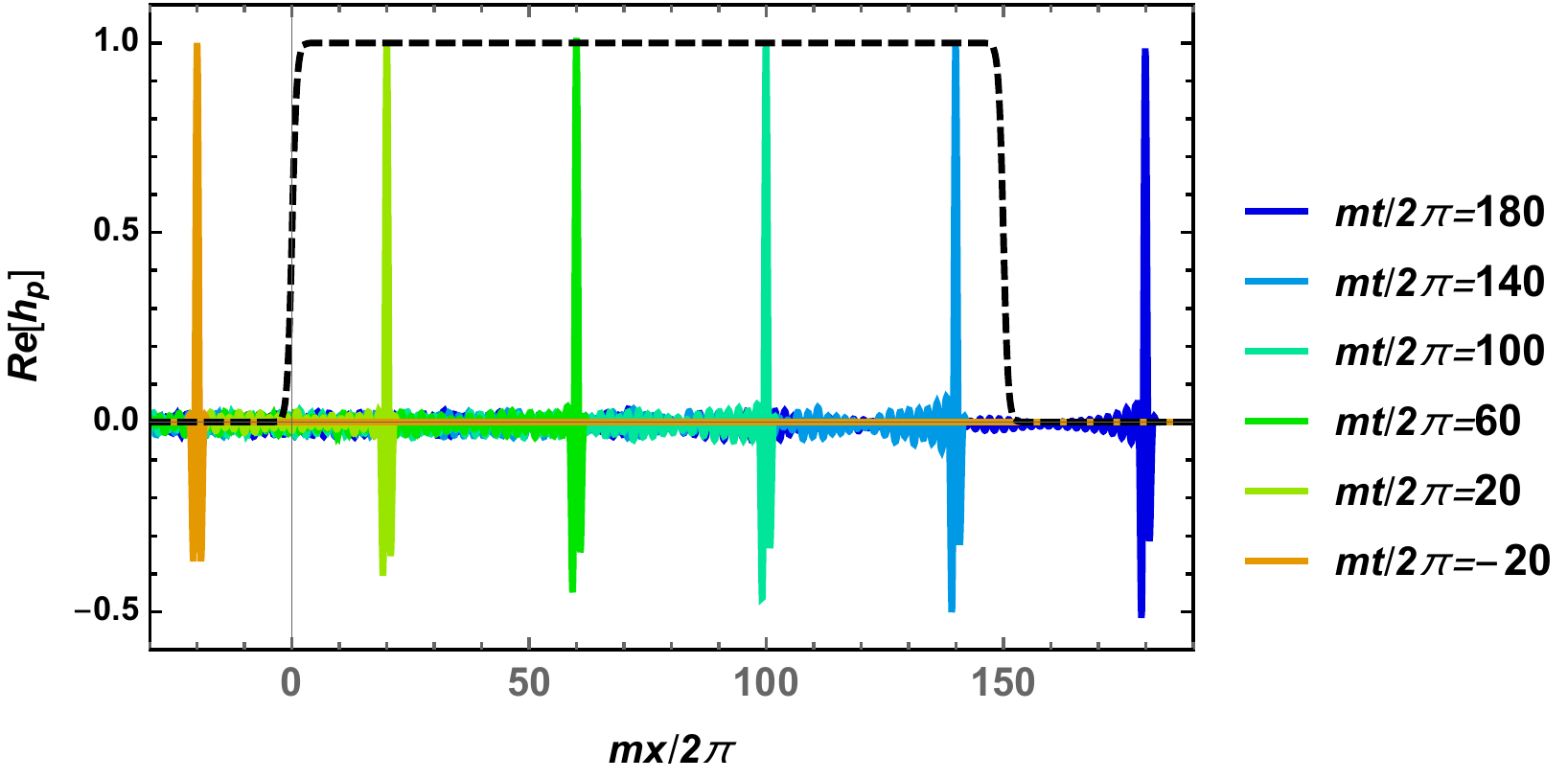}
    \end{center}
  \end{minipage}
  \begin{minipage}{0.45\hsize}
    \begin{center}
      \includegraphics[width=70mm]{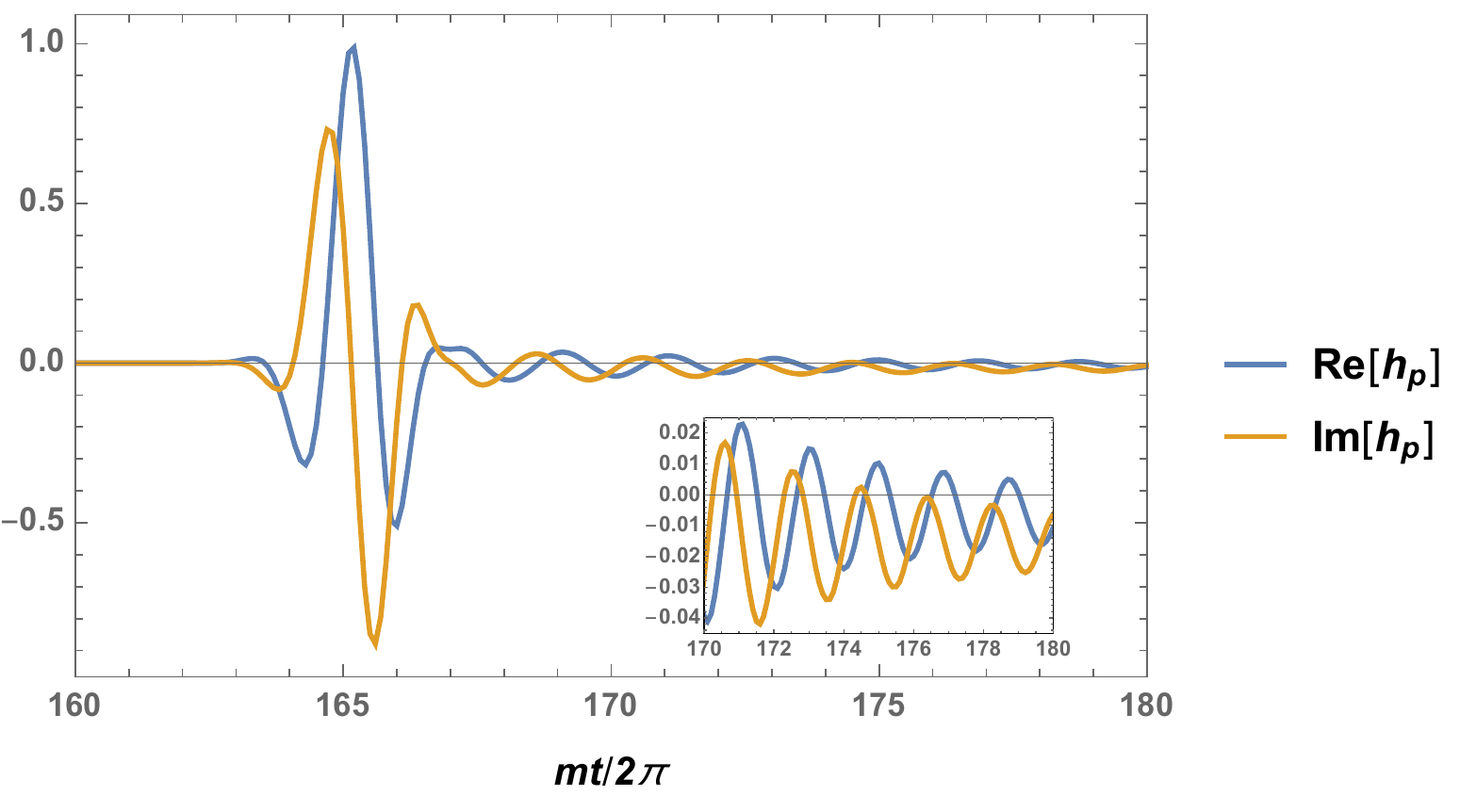}
    \end{center}
  \end{minipage}
  \caption{The left panel shows the real part of the same wave packet as Fig.~\ref{Fig:resonantGW2}. The black dashed line in the left panel shows the axion cloud distribution $|\varepsilon(x)|$. The right panel shows how the GWs would be observed as the function of time outside the axion cloud at $x = 165 \times 2 \pi m^{- 1}$. One can find that in contrast to the coherent case the amplitudes of subsequent waves gradually decrease.}
  \label{Fig:realGW2}
\end{figure}

\begin{figure}[tbp]
  \begin{minipage}{0.45\hsize}
    \begin{center}
      \includegraphics[width=63mm]{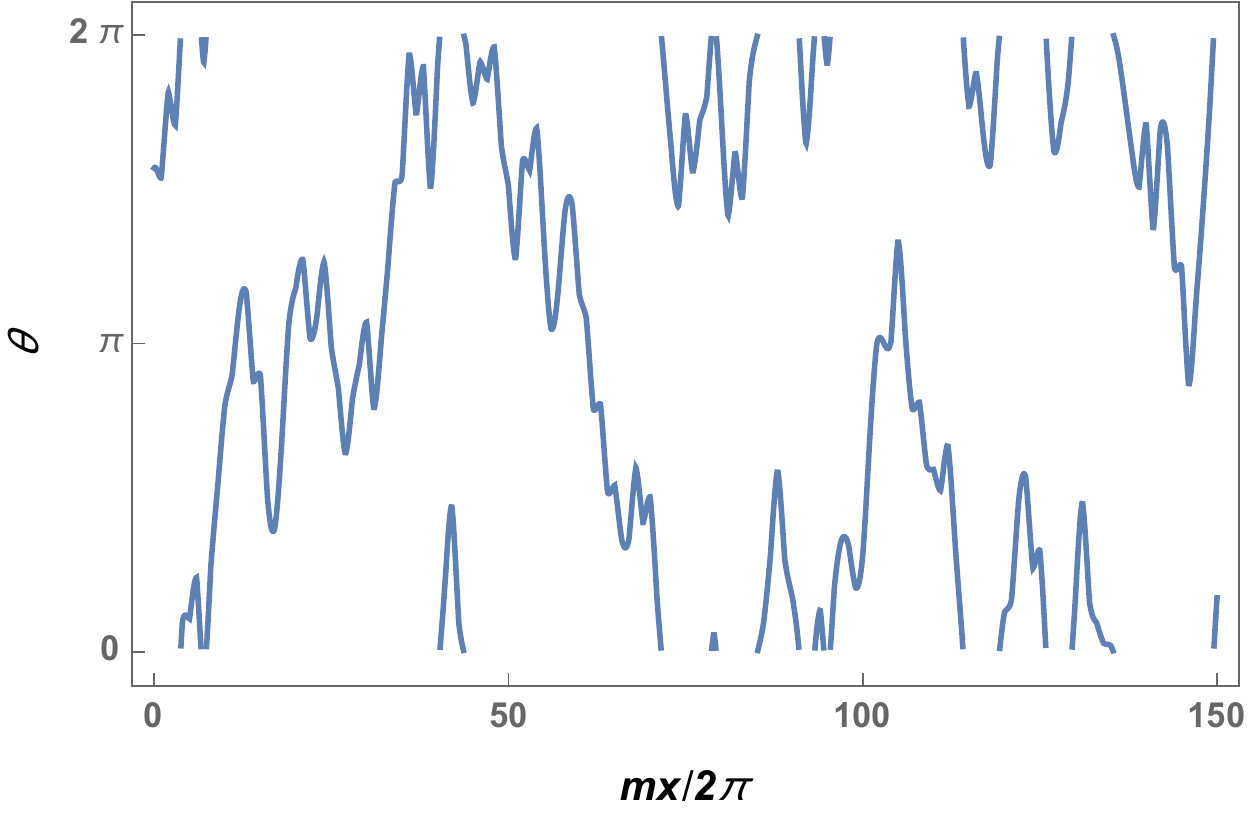}
    \end{center}
  \end{minipage}
  \begin{minipage}{0.45\hsize}
    \begin{center}
      \includegraphics[width=75mm]{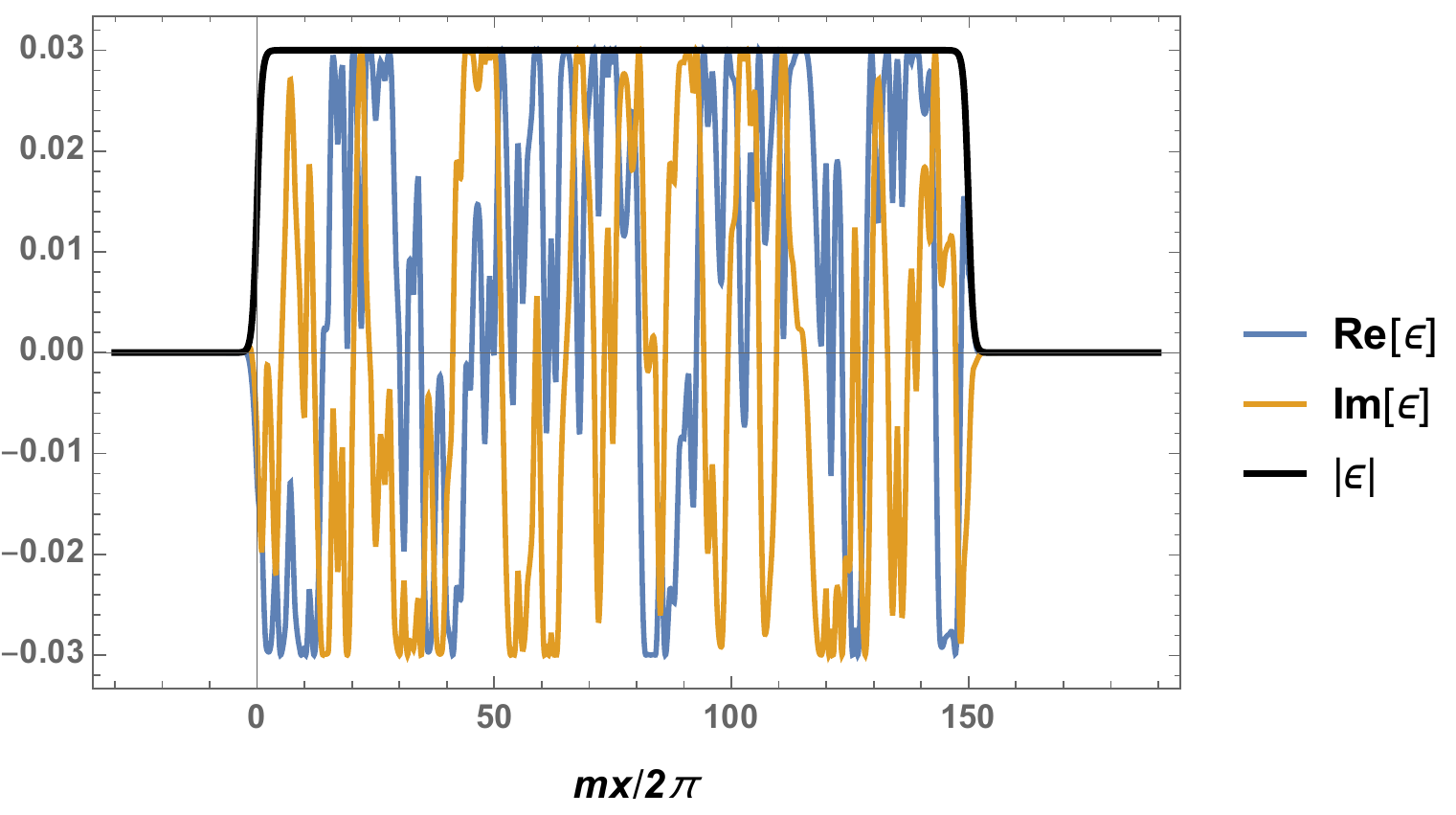}
    \end{center}
  \end{minipage}
  \caption{A realization of $\varepsilon(x)$ used in the numerical calculations for Figs.~\ref{Fig:resonantGW2}--\ref{Fig:realGW2}. Left panel: the phase of the axion $\theta(x)$ generated as a random walk. Right panel: $|\varepsilon(x)|$ (black), $\Re(\varepsilon(x))$ (blue), and $\Im(\varepsilon(x))$ (orange).}
  \label{Fig:deltaphase}
\end{figure}

Figures~\ref{Fig:resonantGW2} and \ref{Fig:realGW2} show the time evolution of the wave packet with $\varepsilon' \neq 0$. We numerically solve Eq.~\eqref{Eq:EoMofh} to obtain the evolution of the wave packet in the incoherent axion cloud. The amplitude of the axion cloud $|\varepsilon(x)|$, its length and the initial condition for the wave packet are the same as in the previous section, but the axion phase, $\theta(x) = {\rm arg}(\varepsilon(x))$, is varied randomly with the specified coherence length as discussed in Sec.~\ref{Sec:Basics_dCS}. $\theta(x)$ shown in Figure~\ref{Fig:deltaphase} is generated by a smooth interpolation of a random walk with the step size of $x$ being $m^{- 1}$ and the root mean square magnitude of one step tuned so as to realize $v = 0.2$. With our choice of parameters for the numerical calculation, the mean value and the standard deviation of $\ln A$ in the limit $ \hdomega \to 0$ are roughly, $\left\langle \ln A(x)\right\rangle \approx 0.5$ and $\sigma_{\ln A} \approx 0.1$, respectively, at $x=150\times 2\pi m^{-1}$.

\section{Cosmological bounds for dCS--axion DM scenarios}
\label{Sec:Condition for the GW resonance}

So far, we have clarified the amplification of GWs passing through an axion cloud in the present universe. In this section, by considering the axion dynamics in the primordial universe, we seek the maximum allowed values of $|\varepsilon|$ and $\mathcal{C}$ at the present time. These parameters characterize the resonant amplification of GWs in the dCS--axion model, as we discussed in the previous sections. Hereafter, we revive the scale factor $a(t)$, and hence the amplitude of the axion's oscillation $|\varphi|$ is time--dependent on the cosmological scale. In the {\it ``standard scenario''} of the axion DM model, the axion starts oscillating when $H(t_{\rm osc}) \sim m$, and it is the dominant component of DM. However, as we will see in Eqs.~\eqref{Eq:Cestimate} and \eqref{eq:finalC}, this prevailing scenario turns out to be too restrictive to satisfy the condition for the interesting amplification to occur, {\it i.e.}, to satisfy $\mathcal{C}\gtrsim 1$. Hence, in the following discussion we assume that the axion can be a sub--dominant DM component and also that the expansion rate $H(t_{\rm osc})$ at the time when the axion oscillations start can be $H(t_{\rm osc}) \ll m$.

If the axion is a minor DM component, the late--time formation of axion clouds would not destroy the successful cosmology scenario. However, even in this case, the late starting time of axion oscillations after the equality time would not lead to the formation of denser compact clouds than the standard scenario. Hence, in a similar way to the formation of ultracompact minihalos~\cite{Ricotti:2009bs}, it is advantageous to consider the gravitational collapse of axion overdensity during the radiation domination in order for a large value of ${\cal C}$ by enhancing the local density of the axion. Therefore, we also assume that the axion starts to oscillate during the radiation dominant era, {\it i.e.}, $H(t_{\rm osc}) > H(t_{\rm eq})$ in the following discussion.

Moreover, it is known that in the dCS gravity one of the two helicity tensor modes becomes ghost--like once its momentum gets larger than a cutoff scale~\cite{Dyda:2012rj}. The condition for the resonant frequency band to be below the cutoff scale can be expressed as
\begin{align}
\label{eq:ghost}
    1 \gtrsim \frac{|\varepsilon(t)|}{4}\,. 
\end{align}
This condition is always satisfied if we avoid the resonant instability due to the spontaneous emission of GWs given in Eq.~\eqref{Eq:condspontaneous}.
Even if the resonant instability occurs, the amplitude of the axion oscillations will drop down to satisfy $|\varepsilon|<\varepsilon_{\rm cri}$ due to the backreaction. By that time, the condition~\eqref{eq:ghost} is automatically satisfied. As we only discuss the evolution of the axion cloud after the instability terminates, we do not have to worry about the constraint \eqref{eq:ghost} any further. 

\subsection{Maximum amplitude of axion oscillations}
\label{Sec.BackReaction}

Here, we discuss the upper bound on the initial amplitude of the axion oscillation while paying attention to the backreaction to the axion cloud due to GW radiation. The resonance occurs when the physical wavenumber of GWs is in the resonant band with a width $m |\varepsilon| / 8$ [see Eq.~\eqref{Eq:resonantband}]:
\begin{equation}
  \frac{m}{2} \left( 1 - \frac{|\varepsilon(t)|}{4} \right) \lesssim \frac{k}{a(t)} \lesssim \frac{m}{2} \left( 1 + \frac{|\varepsilon(t)|}{4} \right).
\end{equation}
As the wavenumber of GWs changes because of the cosmic expansion, the duration $\Delta t$ for which a given GW $k$--mode stays in the resonant band is expressed by   
\begin{align}
  \Delta t \simeq \frac{|\varepsilon(t_k)|}{2H(t_k)} \,,
\end{align}
where $t_k$ denotes the time when the $k$--mode crosses the center of the resonant band, {\it i.e.} $a(t_k) = 2 k m^{- 1}$. Since the maximum growth rate is $m |\varepsilon|/8$, the amplification of each mode is at most
\begin{equation}
  h_k(t_k + \Delta t/2) \simeq e^{N(t_k)} h_k(t_k - \Delta t/2) \,,
\end{equation}
where we define 
\begin{align}
\label{Eq:Noft}
N(t) \equiv \frac{m |\varepsilon(t)|^2}{16 H(t)} = \frac{m |\varepsilon(t_{\rm eq})|^2}{16 H(t_{\rm eq})} \sqrt{\frac{H(t)}{H(t_{\rm eq})}} \,.
\end{align}
For the last equality, we assumed that the backreaction of the GW amplification to the axion is never significant, and thus the axion amplitude decays like $|\varepsilon| \propto |\varphi| \propto a^{-3/2} \propto H^{3/4}$.

Once the instability occurs, the energy of the axion field is transferred to GW background with a sharp peak frequency corresponding to the resonant band at $t_{\rm osc}$. The density parameter of the GWs is estimated as 
\begin{equation}
    \Omega_{\rm GW}(t_0) \simeq \frac{m^2 \varphi^2(t_{\rm osc})}{6 \Mpl^2 H^2(t_{\rm osc})} \Omega_{\rm rad}(t_0) \,.
\end{equation}
The peak frequency at present is 
\begin{equation}
f_{\rm GW} \simeq \frac{m}{2} \left( \frac{a_0}{ a(t_{\rm osc})} \right) \,.
\end{equation}
Here, we should notice that the most of the energy transfer from the axion to GWs occurs instantaneously at $t_{\rm osc}$, because $N(t)$ is largest at $t_{\rm osc}$ and monotonically decreases. Hence the relative width of the peak is $|\varepsilon(t_{\rm osc})|/2$. 

The normalized axion amplitude $|\varepsilon(t)|$, after the primordial GW amplification if any, can be expressed in terms of $N(t)$. Let us introduce a critical time $t_*$ when the decay of the axion amplitude due to the backreaction ceases to be efficient. By definition, $N(t_*)$ should be below a certain critical value of $O(10)$, which would be sufficient to transfer the energy from the axion cloud to GWs:
\begin{equation}
    N_* \equiv N(t_*) < \mathcal{O}(10).
  \label{eq:Nstarbound}
\end{equation}
Using $N_*$, we can describe $\varepsilon(t_{\rm eq})$ as 
\begin{align}
\label{Eq.marginalcond}
    |\varepsilon(t_{\rm eq})| \simeq 
    4 \sqrt{\frac{H(t_{\rm eq})^{3/2} N_*}{m\sqrt{H_*}}} 
    \approx 2 \times 10^{-9} \, \left( \frac{N_*}{10} \right)^{1/2} \left( \frac{m}{10^{-8} \, \eV} \right)^{-1/2} \left(\frac{H(t_{\rm eq})}{H_*}\right)^{1/4} \,, 
\end{align}
with $H_*\equiv H(t_*)$. Since $N_*$ should satisfy the bound \eqref{eq:Nstarbound} and ${H(t_{\rm eq})}/{H_*}\ll 1$, the above equation gives an upper bound on $|\varepsilon(t_{\rm eq})| $ for a given $m$. This condition can be interpreted as a condition on the dCS coupling parameter:
\begin{align}
\label{Eq.marginalcondxi}
\ell_{\rm dCS} 
&\simeq \left( \frac43 \frac{N_*}{\Omega_{\rm a} H(t_{\rm eq}) m^3} \left(\frac{H(t_{\rm eq})}{H_*}\right)^{1/2}\right)^{1/4} \cr
&\approx 3 \times 10^3 \, \Omega_{\rm a}^{-1/4}(t_{\rm eq}) \left( \frac{N_*}{10} \right)^{1/4} \left( \frac{m}{10^{-8} \, \eV} \right)^{-3/4} \left(\frac{H(t_{\rm eq})}{H_*}\right)^{1/8} \, \km \,, 
\end{align}
where $\Omega_{\rm a}(t) = m^2 \varphi^2(t) / (6 \Mpl^2 H^2(t))$. The relation~\eqref{Eq.marginalcondxi} indicates that in order for the axion field to keep the significant energy fraction without decaying due to the GW radiation reaction, the coupling $\ell_{\rm dCS}$ is bounded from above, unless the axion occupies just a tiny fraction of the energy density of the universe. 

If we assume that the dominant component of DM is composed of this axion field, the DM density in our galaxy should be the axion density. Hence, substituting the fiducial values corresponding to our galaxy, we can estimate the value of $\mathcal{C}$ from Eq.~\eqref{Eq:lnAestimate} as
\begin{align}
\label{Eq:Cestimate}
\mathcal{C} &\approx 3 \times 10^{-2} \, \Omega_{\rm a}^{- 1}(t_{\rm eq}) \left( \frac{N_*}{10} \right) \left( \frac{L}{10 \, \kpc} \right) \left( \frac{\rho_{\rm a}}{0.3 \, \GeV/{\rm cm}^3} \right) \left( \frac{v}{7 \times 10^{-4}} \right)^{-1} \left( \frac{H_*}{H(t_{\rm eq})} \right)^{-1/2} \,. 
\end{align}
This clearly shows that the resonant enhancement is inefficient even if the critical time is delayed until the equality time. Therefore, it is difficult for the standard scenario of the axion DM model to realize the efficient amplification of GWs in the present universe.

However, it is too early to conclude that the resonant amplification of gravitational waves in the dCS--axion DM theory never occurs. The above constraint is evaded if the axion is not a dominant component of DM in our universe. In this case, there is no reason to stick to the fiducial values corresponding to the DM cloud associated with our galaxy as a reference. In the next subsection, we discuss this possibility in more detail.

\subsection{Possibility of detectable enhancement of GWs}

For simplicity, we assume that compact axion clouds have been formed via a gravitational contraction of large intrinsic density perturbation. Although the formation process is highly speculative, we expect that this scenario might be realized if the initial spatial variation of the axion field is large. Let us define the critical time $t_\dagger$ when  the collapsing region that forms an axion cloud with the length scale $L$ enters the horizon scale. We also define $H_\dagger = H(t_\dagger)$ and $N_\dagger = N(t_\dagger)$, respectively. At $t=t_\dagger$, the average energy density of the axion field is given by
\begin{align}
\bar{\rho}_{{\rm a} \dagger} &= \frac{4 \Mpl^2 H_\dagger N_\dagger}{m^3 \ell_{\rm dCS}^4} = 3 \Mpl^2 H_\dagger^2 \Omega_{\rm a}(t_\dagger) \,, \\
\label{Eq:OmegaAdagger}
\Omega_{\rm a}(t_\dagger) &= \frac43 N_\dagger \frac{m}{H_\dagger} \left( m \ell_{\rm dCS} \right)^{-4} 
 =\frac43 N_{*} \left(\frac{H_\dagger}{H_*}\right)^{1/2}
 \frac{m}{H_\dagger} \left( m \ell_{\rm dCS} \right)^{-4}\,,
\end{align}
where we have used Eq.~\eqref{Eq:Noft}. Provided that each horizon patch at $t=t_\dagger$ typically collapses into one axion cloud with the size $L$, after the cloud formation, the cloud energy density will be reduced to
\begin{equation}
    \rho_{\rm a} 
    = \frac{\bar{\rho}_{{\rm a} \dagger}}{L^3H_\dagger^3} \, .  
\end{equation}
After its formation, the cloud decouples from the cosmic expansion, and hence the density of the axion cloud is preserved. Therefore, $\varepsilon$ inside the axion cloud at present is given by
\begin{align}
    \varepsilon(t_0) = \sqrt{\frac{16 N_\dagger}{m L^3 H_\dagger^2}} \,.
\end{align}

Assuming that the cloud is virialized, we have an estimate
\begin{equation}
    v \simeq \frac14 \sqrt{\frac{L^2 \rho_{\rm a}}{3 \Mpl^2}} = \frac{\sqrt{N_\dagger m L}}{2 \sqrt{3} ( m \ell_{\rm dCS} )^2 ( L H_\dagger ) } \, ,
    \label{eq:vestimate}
\end{equation}
where we have used $v^2 \simeq M_{\rm cloud} / (8 \pi \Mpl^2 L)$ with $M_{\rm cloud} = 4 \pi (L/2)^3 \rho_{\rm a} / 3$. For consistency, we require $v\ll 1$, which implies the upper bound on $L$:
\begin{equation}
\label{Eq:Lupper} 
     m L \ll m L_{\rm max} \equiv \frac{12}{N_\dagger} (m \ell_{\rm dCS} )^4 ( L H_\dagger )^2 \,, 
\end{equation}
where we treat the non--dimensional combination, $L H_\dagger$, as an independent parameter. We also need to require, at least, $m L >O(1)$ because $L$ must be larger than the (reduced) Compton wavelength, which leads us to the condition 
\begin{equation}
\label{Eq:L0condition}
     m L_{\rm max} \gg 1 \,.
\end{equation}

In Sec.~\ref{Sec:the effect of spatial distribution} we implicitly assumed the cloud size $L$ is larger than the coherence length  $\lambda_{\rm c} \simeq (2 \pi)/m v$. Of course, this condition is not necessarily satisfied. However, if $L < \lambda_{\rm c}$, it is enough to consider the coherent case. Then, the amplification factor, $m \varepsilon L / 8 < m \varepsilon_{\rm cri} L / 8 \sim 1$, cannot be very large. Thus, we focus only on the case satisfying the condition $L > 2 \pi / m v$. Using the expression for $v$ given in Eq.~\eqref{eq:vestimate}, this condition can be rewritten as 
\begin{equation}
  \label{eq:mL1}
  m L > m L_{\rm min} \equiv (4 \pi^2 m L_{\rm max})^{1/3} \gg 1\,,
\end{equation}
where the last inequality holds because of the same reason with Eq.~\eqref{Eq:L0condition}. Since $m L_{\rm max}\gg 1$, the condition~\eqref{eq:mL1} implies
\begin{equation}
\label{Eq:L1vsL0}
     L_{\rm min}\ll L_{\rm max}\, . 
\end{equation}

Furthermore, we also need to require $v^2\lesssim |\varepsilon|$, in order for the discussion in Sec.~\ref{Sec:the effect of spatial distribution} to be valid. Otherwise, the shift due to the axion random motion cannot be neglected, which weakens the GWs amplification. For a moment, let us consider this unfavorable case. The normal dispersion relation of non--relativistic axion, $\omega\simeq m(1+v^2/2)$, implies that the peak of its spectrum at $\omega=m$ is broadened by the velocity dispersion. When $v^2\gtrsim |\varepsilon|$, the part of the spectrum which remains inside the resonant band with the width $O(m |\varepsilon| / 8)$ will be reduced roughly by a factor $|\varepsilon|/v^2$, which is the ratio between the width of the resonant band and the frequency shift due to the axion velocity. As a result, the average amplification $\mathcal{C}$ should be replaced with 
\begin{equation}
    \mathcal{C}_{\rm eff} \equiv \erf \left(\frac{|\varepsilon|}{2 \sqrt{2} v^2}\right) \mathcal{C} \simeq \frac{|\varepsilon|}{\sqrt{2 \pi} v^2} \mathcal{C} \,, 
\end{equation}
where we assume the Gaussian distribution of $\omega$, and $v^2 \gtrsim |\varepsilon|$ is used in the last expression. The transition between two different regimes occurs at $|\varepsilon| = \sqrt{2 \pi} v^2$, {\it i.e.},
\begin{equation}
    m L = m L_{v^2} \equiv m L_{\rm min} \left( \frac{24 (m \ell_{\rm dCS})^4}{\pi^3} \right)^{1/3}\,, 
\end{equation}
and the condition $v^2\lesssim |\varepsilon|$ holds, when $L\lesssim L_{v^2}$. 

As we are interested in placing a tighter constraint on $\ell_{\rm dCS}$, one may think that we should focus on the regime with $m\ell_{\rm dCS} \lesssim 1$. However, in this case we have $L_{v^2} < L_{\rm min} < L$, and hence we should use $\mathcal{C}_{\rm eff}$ for the whole parameter range. When $L$ takes the minimum value within the allowed range, {\it i.e.}, at $L=L_{\rm min}$, $\mathcal{C}_{\rm eff}$ takes the maximum value 
\begin{align}
\mathcal{C}_{\rm eff}(m L_{\rm min}) &= \left( \frac{18 \sqrt{3} N_\dagger^2 \left( m \ell_{\rm dCS} \right)^{10}}{\pi \left(L_{\rm min} H_\dagger\right)^4} \right)^{1/3} \notag\\
&\approx 1 \left( \frac{N_{*}}{10} \right)^{2/3} \left( \frac{m \ell_{\rm dCS}}{0.5} \right)^{10/3} \left( \frac{H_\dagger}{H_*} \right)^{1/3} \left( L_{\rm min} H_\dagger \right)^{-4/3} \,.
\end{align}
We should notice that we naturally expect $L H_\dagger\gtrsim 1$, unless the cloud shrinks after its formation dissipating its energy somehow. Thus, we find that $\mathcal{C}_{\rm eff}$ cannot be large in the regime with $m\ell_{\rm dCS}\lesssim 1$, which means that one cannot expect an observable signature for constraining $\ell_{\rm dCS}$.

Next, we move on to the regime with $m\ell_{\rm dCS}\gtrsim 1$. Now, we have $L_{v^2}>L_{\rm min}$, and the favorable case with $v^2\lesssim |\varepsilon|$ ({\it i.e.}, $L_{v^2}>L$) can be considered, in which we should use the original $\mathcal{C}$. Since $\mathcal C$ is monotonically decreasing in $L$, the maximum value is achieved when $L=L_{\rm min}$. Substituting this value of $L$, we obtain 
\begin{align}
\label{eq:finalC}
    \mathcal{C} &= \frac12 \left( \frac{3 \sqrt{\pi^7 / 2} N_\dagger^2 \left( m \ell_{\rm dCS} \right)^{4}}{\left(L_{\rm min} H_\dagger\right)^4} \right)^{1/3} 
    \approx 10^5 \left( \frac{N_{*}}{10} \right)^{2/3} \left( \frac{m \ell_{\rm dCS}}{10^3} \right)^{4/3} \left( \frac{H_\dagger}{H_*} \right)^{1/3} \left( L_{\rm min} H_\dagger \right)^{-4/3} \,, 
\end{align}
which can be much larger than unity. The fiducial value for $m \ell_{\rm dCS}$ that we adopted here can be realized by choosing $m$ and $\ell_{\rm dCS}$ as 
\begin{align}
m \ell_{\rm dCS} \approx 10^3 \left( \frac{m}{10^{-8} \, \eV} \right) \left( \frac{\ell_{\rm dCS}}{10 \, \km} \right) \,.
\end{align}
$\mathcal{C}$ larger than unity leads to significant amplification, since the amplification is the exponential of $\mathcal{C}$ [see Eq.~\eqref{Eq:AexpC}]. Hence, we conclude that there is a possibility that the frequency dependent amplification of GWs is observed for $m \ell_{\rm dCS} \gtrsim 1$. To realize this possibility, the size of the axion cloud should be close to $L_{\rm min}$, so that the factor $\left( L_{\rm min} H_\dagger \right)^{-4/3}$ in Eq.~\eqref{eq:finalC} should not give a large suppression. Hence, we should set $H_\dagger \simeq L_{\rm min}^{- 1}$, which reads $m / H_\dagger \simeq m L_{\rm min} \gg 1$. In this case, from Eq.~\eqref{Eq:OmegaAdagger}, one can immediately find $\Omega_{\rm a}(t_\dagger) \ll 1$. Therefore, the significant amplification of GWs in the dCS--axion model may take place, if the axion is a minor component of DM and it forms collapsed objects which are relatively compact and dense.

\section{Summary and discussion}
\label{Sec:Summary}

In this paper, we discussed the effects of the parametric resonance in dCS--axion gravity on GWs. At first, focusing on the case where the axion oscillates coherently, we showed that (1) the resonant amplification of GWs is significant in the causal future but there is no amplification in the acausal region, (2) continuous waves subsequent to incident GWs are produced in the narrow resonant frequency band, (3) the axion cloud spontaneously emits GWs if the amplitude of the axion is larger than the critical value determined by the coupling parameter, the mass of the axion, and the size of the axion cloud.

Next, we formulated the ensemble average of the amplification of GWs in the incoherent case. While the resonant amplification is suppressed by $\mathcal{O}(\varepsilon/v)$ compared with the coherent case, we find that we can still expect significant amplification of GWs. We also numerically solved the linearized equations of motion of GWs and compare the results with the analytical expressions.

Furthermore, we discussed how much amplification of GWs is allowed in the present universe by taking account of the cosmic expansion and the backreaction of GW emission of the axion cloud. We found that since resonant amplification is suppressed, it is difficult to test dCS--axion gravity with GW observations in the standard scenario of the axion DM model, in which the axion starts oscillating at $H_{\rm osc} \sim m$ and it is the dominant component of DM. However, there is a possibility that the frequency dependent amplification of GWs would be observed for $m \ell_{\rm dCS} \gtrsim 1$ and $m / H_\dagger \simeq m L_{\rm min} \gg 1$, which requires $\Omega_{\rm a}(t_\dagger) \ll 1$.

In the above analysis, we focus only on the dominant effect of the parametric resonance of GWs by linearization and we neglect the effects of the environment, such as the host galaxy if the DM halo consists of the axion, and the higher--order terms (see also Refs.~\cite{Marsh:2015xka,Kolb:1993zz,Schive:2014dra}). We leave it for future work to take into account such effects and to consider the detailed density profiles of axion clouds.

Although the stringent constraint from NICER $\ell_{\rm dCS} \lesssim 10 \, \rm{km}$ requires the resonant frequency ranges to be $\gtrsim 2 \times 10^3 \, {\rm Hz}$ which is slightly higher than the best sensitivity band  of the current ground--based laser interferometers, we might have a chance to prove with the future high--frequency GW detectors~\cite{Akutsu:2008qv,Cruise:2012zz,Ito:2019wcb,Chen:2020ler}. Regarding the possible source of such gravitational waves with very high frequency $f \sim 10^6 - 10^9 \, {\rm Hz}$ ({\it i.e.}, $m \sim 10^{-8} - 10^{-5} \, \eV$ for the peak resonance), we may expect the cosmological origin provided at the end of inflation or just after inflation ({\it e.g.}, see Refs.~\cite{Khlebnikov:1997di,Ito:2016aai,Adshead:2018doq}). Moreover, primordial black holes evaporation may also be an interesting target~\cite{BisnovatyiKogan:2004bk}. Because of the interesting feature of the parametric resonance that the continuous waves after incident GWs are produced in the narrow resonant frequency band, to probe/constrain the resonance, it might be suitable to perform a kind of residual test that subtracts the GR best--fitting waveform from data to seek subsequent resonant waves.

\section*{Acknowledgments}

We wish to acknowledge the Yukawa Institute for Theoretical Physics at Kyoto University, where a part of this work was initiated during ``The 3rd Workshop on Gravity and Cosmology by Young Researchers'' (YITP-W-18-15). This work was supported by JSPS KAKENHI Grant Number JP17H06358 (and also JP17H06357), A01: {\it Testing gravity theories using GWs}, as a part of the innovative research area, ``GW physics and astronomy: Genesis''. T.F. acknowledges support from JSPS KAKENHI Grant No. JP18K13537. T.T. acknowledges support from JSPS KAKENHI Grant No. JP20K03928. I.O. acknowledges supported from JSPS Overseas Research Fellowship.

\appendix

\section{Calculation of the wave packet}
\label{sec:Calculation of the wave packet}

Here, we evaluate the integral in Eq.~\eqref{Eq:wavepacket} with the method of the steepest decent. According to this method, the complex integral of $e^{g(z)}$ along the path of the steepest descent passing through a saddle point $z_{\rm sp}$ is approximately obtained as
\begin{equation} 
\label{Eq:saddleformula}
  \int \dd z\, e^{g(z)} \approx \left(\frac{2\pi}{|g''(z_{\rm sp})|}\right)^{1/2}\,e^{g(z_{\rm sp})}\,.
\end{equation}
The saddle points of the exponent $f(\domega)$ defined in Eq.~\eqref{Eq:phaseofpacket} is given by the root of
\begin{equation}
  \label{Eq:dfdomega}
  \frac{\partial f}{\partial \domega} = - i t + i x \left( 1 + \frac{|\varepsilon|^2}{64} \frac{m^2}{\domega^2} \right)^{-1/2} - \frac{\domega}{K^2} = 0 \,.
\end{equation}
This equation has three different roots. One is located in the vicinity of the solution of the vanishing $\varepsilon$ case,
\begin{align}
  \label{Eq:packetdomega}
  \domega_0 = - i ( t - x ) K^2 + \frac{i m^2 x |\varepsilon|^2}{128 K^2 ( t - x )^2} + \mathcal{O}(\varepsilon|^4).
\end{align}
Furthermore, as we are interested in the asymptotic region where both $t$ and $x$ are large, neglecting the last term of R.H.S. in Eq.~\eqref{Eq:dfdomega} we obtain the other two stationary points as
\begin{equation}
  \label{Eq:instabilitydomega}
  \domega_s^{\pm} = \pm i \frac{m |\varepsilon| t}{8 \sqrt{t^2 - x^2}} \,.
\end{equation}
It is interesting to note that $\delta \omega_s^\pm$ is pure imaginary for $x<t$ while real for $x>t$. 

\begin{figure}[tbp]
  \begin{minipage}{0.45\hsize}
    \begin{center}
      \includegraphics[width=70mm]{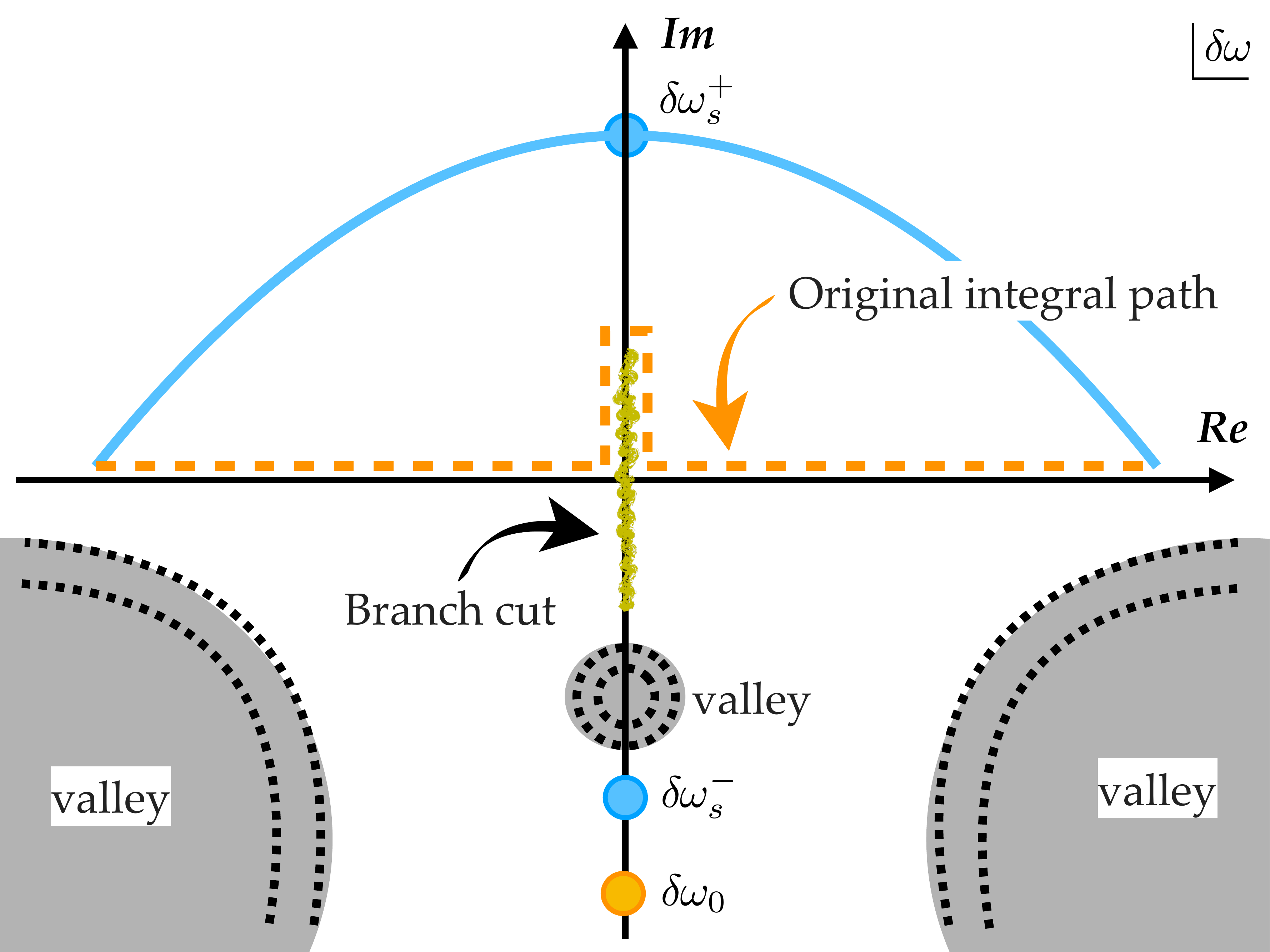}
    \end{center}
  \end{minipage}
  \begin{minipage}{0.45\hsize}
    \begin{center}
      \includegraphics[width=70mm]{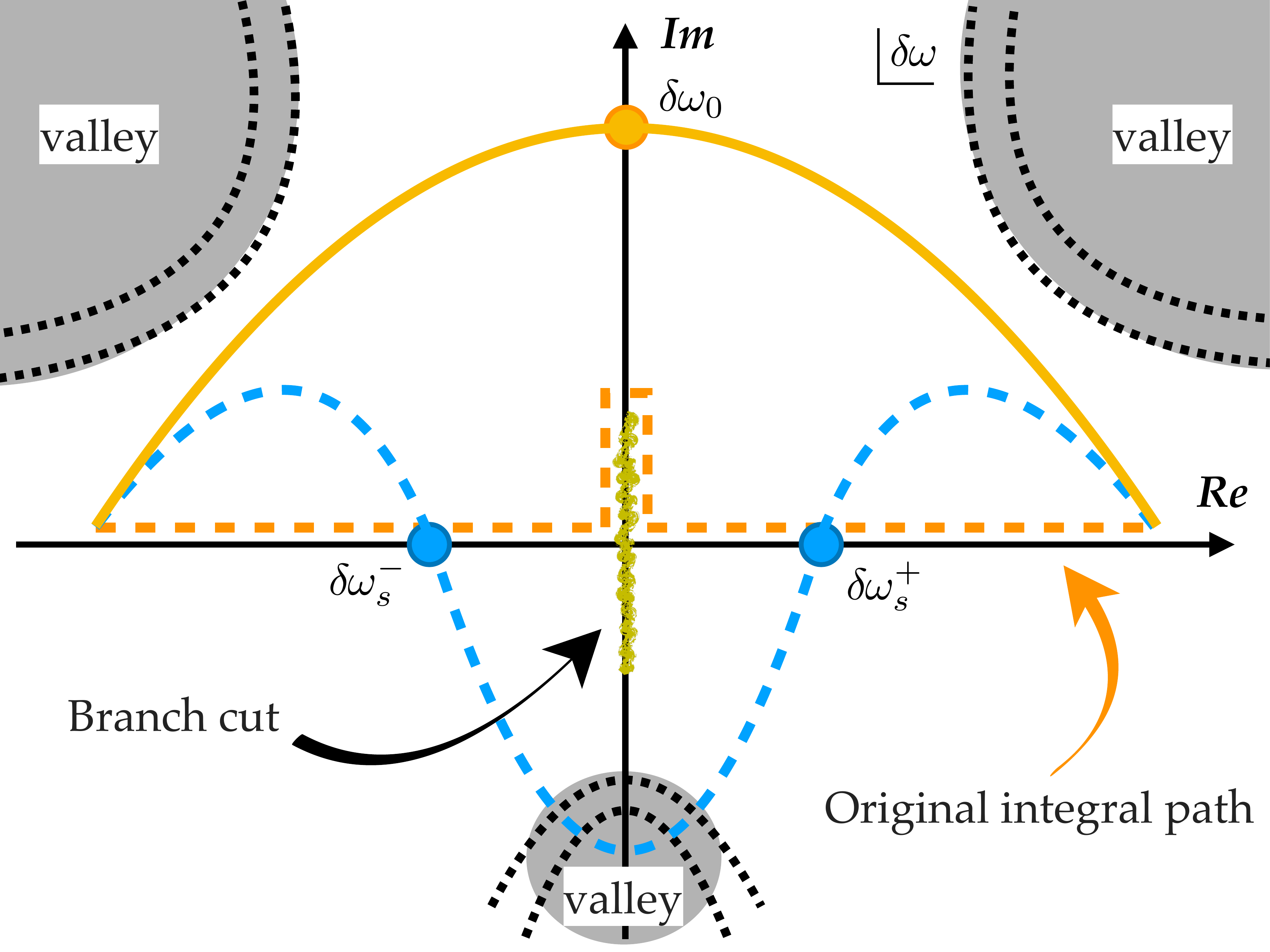}
    \end{center}
  \end{minipage}
  \caption{Schematic pictures of the saddle points and integration paths in the causal future $t>x$ ({\it left panel}) and the acausal region $x>t$ ({\it right panel}). The black dashed lines represent the contours of the real part of $f(\domega)$ which have several valleys. The original integration path (orange dashed lines) in Eq.~\eqref{Eq:wavepacket} should pass above the branch cut for our initial condition. Therefore, one obtains the same result by integrating it on a path along the steepest descent line through the saddle point above the brunch cut (solid lines). For $x>t$, the steepest descent line passing through $\domega_s^\pm$ goes below the brunch cut due to the configuration of $f(\domega)$ (blue dashed line).}
  \label{Fig:saddle}
\end{figure}

\begin{figure}[tbp]
  \begin{center}
    \includegraphics[width=100mm]{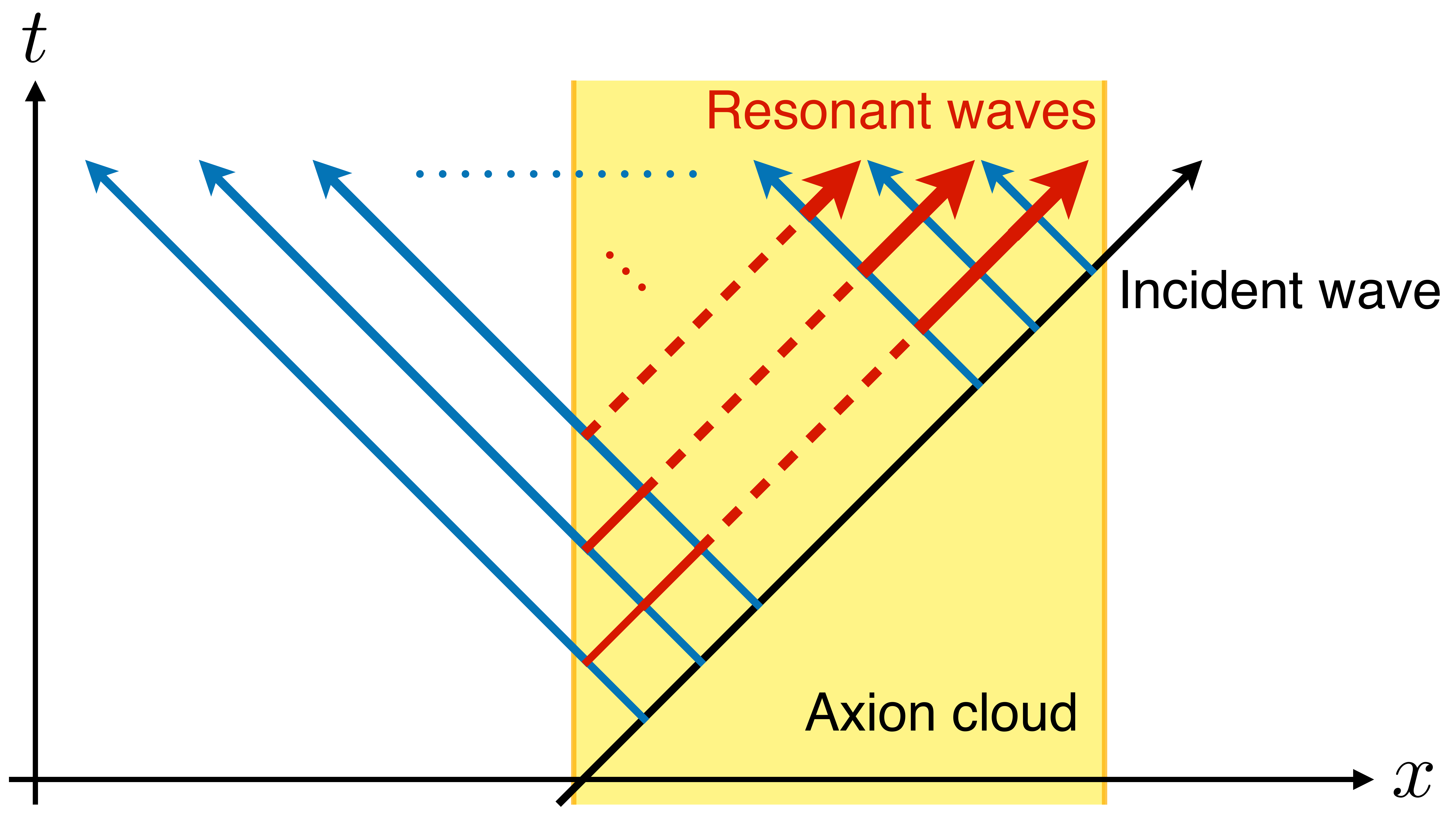}
  \end{center}
  \caption{A schematic figure of the production of the resonant waves. In the absence of incident left--going waves, the resonant waves (red) are produced after the left--going waves (blue) are induced by the interaction between the incident wave (black) and the axion in the axion cloud (yellow box).}
  \label{Fig:schematic}
\end{figure}

Now, we perform the integral in Eq.~\eqref{Eq:wavepacket} on the complex plane of $\domega$. First, let us consider the case of the causal future, $t>x$. $f(\domega)$ has a branch cut on the imaginary axis of $\domega$ between $\domega=-i m|\varepsilon|/8$ and $\domega=+i m|\varepsilon|/8$. Since $e^{f(\domega)}$ is required to vanish in the past infinity $t\to -\infty$ by our initial condition, the original integral path passes over the branch cut. As shown in the left panel of Fig.~\ref{Fig:saddle}, the saddle point $\domega_s^+$ is picked in this case. Then, using Eq.~\eqref{Eq:saddleformula}, we obtain
\begin{equation}
h_{\rm packet}(t>x) \approx \left( 1 + \frac{8(t^2-x^2)^{3/2}}{m|\varepsilon|x^2} K^2 \right)^{-\frac{1}{2}} \exp\left[\frac{m |\varepsilon|}{8} \sqrt{t^2 - x^2}\right] \,.
\end{equation}
It implies an exponential growth $h_{\rm packet}\sim e^{m|\varepsilon| t/8}$ in the late time  $t\gg x$, which reproduces Eq.~\eqref{eq:resonantGrowth}. In the acausal region  $x>t$, however, the saddle point appearing above the brunch cut is $\domega_0$, and the other two $\domega_s^\pm$ brought by the oscillating axion are irrelevant to the integral, as illustrated in the right panel of Fig.~\ref{Fig:saddle}. Thus, one finds 
\begin{equation}
h_{\rm packet}(x>t) \approx \left(1 + \frac{m^2 |\varepsilon|^2 x}{64 K^4 ( x - t )^3} \right)^{-\frac{1}{2}}
\exp\left[-\frac{1}{2} K^2(x - t)^2+\frac{m^2|\varepsilon|^2 x}{128 K^2 ( x - t )}\right]\,.
\end{equation}
This result coincides with the normal wave packet up to the $\mathcal{O}(|\varepsilon|^2)$ correction and a resonant growth is absent. This result can be intuitively understood that in the absence of incident left--going waves the resonant waves are produced after the left--going waves are induced by the interaction between the incident wave and the axion (see Figure~\ref{Fig:schematic}).

It should be noted that our previous saddle points,  $\domega_0$ and $\domega_s^\pm$, are invalid for $x = t$. Substituting $x = t$ into Eq.~\eqref{Eq:dfdomega} and solving it at the leading order of $|\varepsilon|$, we obtain
\begin{equation}
  \domega =  i \left( \frac{t m^2 |\varepsilon|^2 K^2}{128} \right)^{1/3} \,.
\end{equation}
This time, a non--trivial solution is found on the imaginary axes above the brunch cut. Substituting back this solution into the integral expression for $h_{\rm packet}$, we obtain
\begin{align}
 h_{\rm packet}(t = x) \approx \left( 3 + \left( \frac{27 m^2 |\varepsilon|^2}{16 K^4 t^2} \right)^{1/3} \right)^{-1/2} \exp \left[ \frac{3}{32} \left( \frac{m^2 |\varepsilon|^2 t}{2 K} \right)^{2/3} \right] \,.
\end{align}
This means that the amplification of the center of the wave packet is less efficient compared to the causal future region $t>x$. 

\section{Averaged solution in the incoherent case}
\label{Sec:AveragedSolution}

Using Eqs.~\eqref{Eq:solX} and \eqref{Eq:gaussepsilon}, we can estimate the ensemble average of $|X^2|$ as
\begin{align}
\langle |X^2(x)| \rangle 
&= \frac{m^2}{64} |\varepsilon(x)|^2 \int_x^{x_{\rm end}} \dd y \int_x^{x_{\rm end}} \dd z \, \frac{m^2}{64} \varepsilon(y) \varepsilon^*(z) e^{- 2 i \domega ( y - z )} \notag\\
&\simeq \frac{m^2}{64} |\varepsilon(x)|^2 \left[ \frac{\sqrt{\pi/2}}{64} m^2 \lambda_{\rm c} e^{- \hdomega^2} \left[ 1 + \erf \left( i \hdomega \right) \right] |\bar \varepsilon|^2 ( x_{\rm end} - x ) \right] \notag\\
&= \frac{m^2}{64} |\varepsilon(x)|^2 \left( \mathcal{C} - \left\langle \ln A(x) \right\rangle \right) \,.
\end{align}
Therefore, in the early time when $\left\langle \ln A(x) \right\rangle$ is negligible, the ratio between the $X^2$--term and the last term in Eq.~\eqref{Eq:approxEqX2}, {\it i.e.}, $m^2|\varepsilon|^2/64$, is $\mathcal{O}(\mathcal{C})$. Thus, one might think it unjustified to neglect the $X^2$--term when $\mathcal{C} \gtrsim 1$. However, even in this case, the contribution from the $X^2$--term is subdominant for the ensemble average of the solution~\eqref{Eq:MeanX}, because of the rapidly varying phase of the $X^2$--term. To make this suppression clear, let us solve the equation of motion without neglecting the $X^2$--term. 

The equation of motion~\eqref{Eq:masterEqX} can be written as
\begin{align}
A'' + \left( 2 i \domega - \frac{\varepsilon'}{\varepsilon} \right) A' + \frac{m^2 |\varepsilon|^2}{64} A = 0 \,.
\end{align}
Reconsidering this equation as a first--order differential equation for $A'$, a formal solution is obtained as
\begin{align}
  \label{Eq:dAgen}
  A'(x) = C_1 \varepsilon(x) e^{- 2 i \domega x} + \frac{m^2}{64} \int_x^{x_{\rm end}} \dd y A(y) \varepsilon(x) \varepsilon^*(y) e^{- 2 i \domega ( x - y )} \,,
\end{align}
where $C_1$ is a constant of integration. As we did to obtain Eq.~\eqref{Eq:incoherentB}, the right--boundary condition $B(x_{\rm end}) = 0$ leads to 
\begin{align}
  C_1 = 0 \,.
\end{align}
Now, in order to obtain the trend of the average of $A$, we assume that $\varepsilon$ is sufficiently small. Thus, using $A(y) = A(x) + \int_x^y \dd z A'(z)$ and Eq.~\eqref{Eq:dAgen}, iteratively, one finds
\begin{align}
A'(x) &= \frac{m^2}{64} A(x) \int_x^{x_{\rm end}} \dd y \left[ 1 + \frac{m^2}{64} \int_x^y \dd z \int_z^{x_{\rm end}} \dd u \, \varepsilon(z) \varepsilon^*(u) e^{- 2 i \domega ( z - u )} + \mathcal{O}\left( \varepsilon^4 \right) \right] \notag\\
&~~~ \times \varepsilon(x) \varepsilon^*(y) e^{- 2 i \domega ( x - y )} \,.
\end{align}
Therefore, we find
\begin{align}
\label{Eq:iterativeX}
X(x) &= \frac{A'(x)}{A(x)} \notag\\
&\simeq \frac{m^2}{64} \int_x^{x_{\rm end}} \dd y \, \varepsilon(x) \varepsilon^*(y) e^{- 2 i \domega ( x - y )} \notag\\
&~~~ + \frac{m^4}{4096} \int_x^{x_{\rm end}} \dd z \int_z^{x_{\rm end}} \dd y \int_z^{x_{\rm end}} \dd u \, \varepsilon(x) \varepsilon^*(y) \varepsilon(z) \varepsilon^*(u) e^{- 2 i \domega ( x + z - y - u )} \,.
\end{align}
In the last line, we swapped the order of integrations with respect to $y$ and $z$. One can observe the above iterative solution~\eqref{Eq:iterativeX} satisfies the equation of motion~\eqref{Eq:masterEqX} up to $\mathcal{O}(\varepsilon^2)$. The first term is identical to Eq.~\eqref{Eq:solX}. As we have
\begin{align}
\langle \varepsilon(x) \varepsilon^*(y) \varepsilon(z) \varepsilon^*(u) \rangle &= \langle \varepsilon(x) \varepsilon^*(y) \rangle \langle \varepsilon(z) \varepsilon^*(u) \rangle + \langle \varepsilon(x) \varepsilon^*(u) \rangle \langle \varepsilon(z) \varepsilon^*(y) \rangle \,,
\end{align}
for Gaussian distribution, the ensemble average of $X(x)$ is evaluated as 
\begin{align}
\label{Eq:iterativeXave}
\langle X(x) \rangle &\simeq \frac{m^2 |\bar{\varepsilon}|^2}{64} \sqrt{\frac{\pi}{2}} \lambda_{\rm c} e^{- \hdomega^2} \left[ 1 + \erf \left( i \hdomega \right) \right] \notag\\
&~~~ \times \left[ 1 + \frac{m^2 |\bar{\varepsilon}|^2 \lambda_{\rm c}^2}{32} \left( 1 + i \sqrt{\pi} e^{- \hdomega^2} \hdomega \left\{ 1 + \erf \left( i \hdomega \right) \right\} \right) \right] \,, 
\end{align}
where we have assumed $x_{\rm end} - x \gg \lambda_{\rm c}$. The second term in the last square brackets in Eq.~\eqref{Eq:iterativeXave} can be neglected as long as 
\begin{align}
m |\bar{\varepsilon}| \lambda_{\rm c} \ll 1 \,,
\end{align}
which is consistent with no spontaneous emission of GWs, and $A' \simeq 0$ in each coherent patch. Therefore, we can integrate the above equation to obtain
\begin{align}
  \langle \ln A(x) \rangle \simeq \frac{\sqrt{\pi/2}}{64} m^2 \lambda_{\rm c} e^{- \hdomega^2} |\bar \varepsilon|^2 \left[ 1 + \erf \left( i \hdomega \right) \right] ( x - x_0 ) \,.
\end{align}
This is the same with Eq.~\eqref{Eq:MeanX}. Therefore, it is justified to neglect the $X^2$--term in the equation of motion in order to evaluate the averaged solution of the left--going wave, which describes the efficiency of the resonant amplification.

\bibliography{Ref}

\end{document}